\documentclass{jfm}
\usepackage{graphicx}
\usepackage{pgfplots}
\usepackage{subfig}
\newcommand{\changes}[1]{{\color{black} #1}} %Review changes round 1
\newcommand{\cchanges}[1]{{\color{black} #1}} %Review changes round 2

\usepackage{amsmath}

\shorttitle{Capillary waves with surface viscosity}
\shortauthor{L. Shen, F. Denner, N. Morgan, B. van Wachem and D. Dini}

\title{Capillary waves with surface viscosity}

\author{Li Shen\aff{1}
  \corresp{\email{l.shen14@imperial.ac.uk}},
  Fabian Denner\aff{1},
  Neal Morgan\aff{2},
  Berend van Wachem\aff{1,3}
 \and Daniele Dini\aff{1}}

\affiliation{\aff{1}Department of Mechanical Engineering, Imperial College London, London,
SW7 2AZ, UK
\aff{2}Shell Global Solutions Ltd, Shell Centre, York Road, London, SE1 7NA, UK
\aff{3}Lehrstuhl f\"{u}r Mechanische Verfahrenstechnik, Otto-von-Guericke-Universit\"{a}t Magdeburg, Universit\"atsplatz 2, 39106 Magdeburg, Germany}

\begin{document}
\maketitle

\begin{abstract}
Experiments over the last 50 years have suggested a tentative correlation between the surface (shear) viscosity and the stability of a foam or emulsion. We examine this link theoretically using small-amplitude capillary waves \changes{in the presence of a surfactant solution of dilute concentrations} where the associated Marangoni and surface viscosity effects are modelled via the Boussinesq-Scriven formulation. \cchanges{The resulting integro-differential initial value problem is solved analytically} and surface viscosity is found to contribute an overall damping effect on the amplitude of the capillary wave with \changes{varying degrees depending on the lengthscale of the system. \cchanges{Numerically,} we find the critical damping wavelength to increase for increasing surface concentration but the rate of increase remains different for \cchanges{both the }surface viscosity and the Marangoni effect. } %This has direct consequences to the stochastic processes of thin film rupture, the onset of which is induced by the roughness of the surface caused by the capillary wave. Henceforth, the dynamics of the capillary wave can, for instance, provide key information to shed light on the mechanism leading to foam formation and collapse. 

\end{abstract}

\section{Introduction}\label{sec:1intro}

Capillary waves on a viscous fluid interface have recently been observed \citep{Aarts2004} to induce the spontaneous breakup of a thin liquid film and controls the inherent stochastic process of the sub-micron rupture event. Unlike gravity waves, these capillary waves have a short wavelength where the restoring force of surface tension dominates over the influence of gravity and can be found in the study of small-lengthscale interfacial phenomena; for instance, thin liquid films \citep{Scheludko1967} and droplet coalescence \citep{Blanchette2006}. It is apparent that variations in surface tension can have dramatic knock-on effects on the dynamics of the capillary waves, with applications found both in surface chemistry \citep{Edwards1991} and interfacial fluid dynamics \citep{Levich1969}.

Surface-active materials, or surfactants, often lead to the formation
of foams and emulsions by lowering the surface tension of a liquid
interface \citep{Batchelor2003,Levich1969,Edwards1991}. Gradients
of surfactant concentration (and therefore the surface tension coefficient) caused
by dilatational deformations induce the Marangoni stress, which acts
to oppose the changes in surface area and slows down the drainage
and rupture processes of a thin liquid film. Moreover, the two-dimensional surfactant monolayer displays the rheological
response whereby shearing deformations can introduce an extra surface
shear viscosity. In addition, a source of surface dilatational viscosity
can result from the inherent compressibility of the two-dimensional
surfactant monolayer \citep{Zell2014}; in direct contrast with the incompressible Newtonian bulk
fluid which can be characterised entirely by a single viscosity parameter
$\mu$. Furthermore, the dissipative nature of the surfactant adsorption-desorption
kinetic process can also contribute towards the effective surface
dilatational viscosity \citep{Lucassen1966}. With multiple sources
of surface viscosities, we henceforth denote the effective surface
dilatational and shear viscosity by $\mu_{d}$ and $\mu_{s}$, respectively.
Finally, we note that the magnitude of $\mu_{d}$ and $\mu_{s}$ need
not be comparable \citep{Djabbarah1982}, since they are each responsible
for different physical processes.  The physical manifestation of surface viscosity and its measurement
remain controversial and subtle; for decades of literature cannot
agree on measurements of $\mu_{s}$ and $\mu_{d}$, the chief
difficulty lies with the fact that \cchanges{not only are surface viscosity and Marangoni effects intimately intwined  \citep{Levich1969,Scheid2010, Langevin2014},} experiments give
the total characteristics for both the surface and bulk phases simultaneously
and it is not trivial to extract the surface information a-priori
of the establishment of a particular surface model.  

For insoluble surface-active (surfactants) solutions, the intrinsic surface shear viscosity is clearly defined. However, for soluble surfactant solutions, in particular sodium dodecyl sulfate (SDS), the presence of a 3-dimensional sublayer adjacent to the surface alters the rates of surface deformation \citep{Stevenson2005}, which may explain the numerous inconsistencies in the reported literature of the magnitudes of surface shear viscosity of surfactants. Some progress has been made recently, namely by the experimental work of \citet{Zell2014}, in which the use of microbutton surface rheometry appears to yield relatively unambiguous measurements of the surface shear viscosity $\mu_s$ of SDS. They report an upper bound of $\mu_s\sim O(10^{-8}\mathrm{Nsm}^{-1})$, which suggests that surface shear viscosity need not be the dominant surface phenomena and that Marangoni effects and surface dilational viscosity may also be in effect. 

In insoluble surfactant solutions, the surface shear viscosity is often much higher than $O(10^{-8}\mathrm{Nsm}^{-1})$, in particular, in the case of 1-eicosanol, it is found \citep{Zell2014,Gavranovic2006} to be at least $10^3$-$10^4$ times higher than that of soluble SDS solutions. Moreover, recent numerical \citep{Gounley2016} and experimental studies conclude that surface viscosity effects in insoluble surfactants can give rise to noticeable behaviours on the resulting
dynamics,  which cannot otherwise be fully
understood if we considered the Marangoni effect alone \citep{Ponce-Torres2017}. In this paper, we shall investigate both the effects of Marangoni and surface viscosity in insoluble surfactant solutions \changes{with a particular focus on the dynamics of very thin films with capillary waves close to critical damping. For such a thin film geometry of high wavenumber, we may consider a two-dimensional flow structure as well as a low Reynolds number under the Stokes' limit. Foreshadowed by the previous numerical work \citep{Gounley2016,Ponce-Torres2017}, we anticipate a similar importance of the Marangoni and surface viscosity effects on the capillary wave in the two-dimensional thin film case under the Stokes' limit.}

In \S\ref{sec:2surf_vis}-\ref{sec:4solution}, we extend the previous work of \citet{Prosperetti1976} and \citet{Shen2017} to incorporate both surface shear and dilatational viscosity to the leading-order, as described by the Boussinesq-Scriven model, into the
dynamics of small-amplitude capillary waves. We delineate the effects
of the convective-diffusive Marangoni stresses with surface viscosity
effects in \S 5. \changes{In \S 6, we obtained an analytical form of the critical damping wavelength for the clean case considering only the bulk fluid viscosity. In \S7, we outline a numerical method to calculate the damping ratio of a general higher-ordered system and construct a minimal pole matrix to encode the information of the poles of the system with significant residue. Under this approach, we identify the transition point of the wave from an underdamped to overdamped state of a general system and obtain numerically the correction to this critical wavelength by surface viscosity and Marangoni effects.} The article is concluded in \S8. 
\section{Boussinesq-Scriven surface viscosity}\label{sec:2surf_vis}
Under the Boussinesq-Scriven model of surface viscosity \citep{Scriven1960,Aris1963,Slattery2007}, the surface stress boundary conditions at the interface
between two Newtonian fluids can be written as \changes{
\begin{equation}
[\mathbf{n}\cdot\mathbf{T}]=\boldsymbol{\nabla}_{s}\cdot\boldsymbol{\sigma}_{s},
\end{equation}
where $\mathbf{T}$ is the viscous stress tensor, $\boldsymbol{\nabla}_{s}=\mathbf{P}\cdot\boldsymbol{\nabla}$ is the surface gradient operator for the projection tensor $\mathbf{P}=\mathbf{I}-\mathbf{nn}^\mathrm{T}$ with normal vector $\mathbf{n}$, $[\cdot]$ denotes the jump
in magnitude across the interface and $\boldsymbol{\sigma}_{s}$ is the surface viscous stress tensor defined by 
\begin{equation}
\boldsymbol{\sigma}_{s}=\sigma \mathbf{P}+(\mu_{d}-\mu_{s})(\boldsymbol{\nabla}_{s}\cdot\mathbf{u}_s)\mathbf{P}+2\mu_{s}\mathbf{D}_{s},
\end{equation}
and
\begin{equation}
\mathbf{D}_{s}=\frac{1}{2}\left(\mathbf{P}:\boldsymbol{\nabla}_{s}\mathbf{u}_{s}+(\boldsymbol{\nabla}\mathbf{u}_s)^{\mathrm{T}}:\mathbf{P}\right)
\end{equation}
is the surface rate of deformation tensor. The divergence of $\boldsymbol{\sigma}_{s}$ may be written \citep{Scriven1960} in the form  
\begin{eqnarray}
[\mathbf{n}\cdot\mathbf{T}] & =& \boldsymbol{\nabla}_{s}\sigma+(\mu_{d}+\mu_{s})\boldsymbol{\nabla}_{s}(\boldsymbol{\nabla}_{s}\cdot\mathbf{u}_s)\nonumber\\
& +&  \mu_{s}\left[2K\mathbf{u}_{s}+\mathbf{n}\times\boldsymbol{\nabla}_{s}(\mathbf{n}\cdot\boldsymbol{\nabla}_{s}\times\mathbf{u}_s)+2(\mathbf{n}\times\boldsymbol{\nabla}_{s}\mathbf{n}\times\mathbf{n})\cdot\boldsymbol{\nabla}_{s}(\mathbf{u}\cdot\mathbf{n})\right]\nonumber \\
&   +& \mathbf{n}\left[2H\sigma+2H(\mu_{d}+\mu_{s})\boldsymbol{\nabla}_{s}\cdot\mathbf{u}_s-2\mu_{s}(\mathbf{n}\times\boldsymbol{\nabla}_{s}\mathbf{n}\times\mathbf{n}):\boldsymbol{\nabla}_{s}\mathbf{u}_s\right],\label{eq:SCRIVEN}
\end{eqnarray} where  
\begin{eqnarray}
2H & = & -\boldsymbol{\nabla}_{s}\cdot\mathbf{n},\\
2K & = & -(\mathbf{n}\times\boldsymbol{\nabla}_{s}\mathbf{n}\times\mathbf{n}):\boldsymbol{\nabla}_{s}\mathbf{n},
\end{eqnarray}
are the mean and Gaussian  curvatures of a surface, respectively, and $\sigma$ is the surface tension coefficient.} Neglecting higher-order terms, the leading-order surface stress boundary condition in the context of small-amplitude capillary waves takes the reduced form 
\begin{equation}
[\mathbf{n}\cdot\mathbf{T}]=\boldsymbol{\nabla}_{s}\tilde{\sigma}+2H\tilde{\sigma}\mathbf{n},\label{eq:LIN-STRESS}
\end{equation}
where $\tilde{\sigma}$ is the surface tension augmented with the leading-order surface viscosity contribution given by 
\begin{equation}
	\tilde{\sigma}=\sigma+(\mu_{d}+\mu_{s})\boldsymbol{\nabla}_{s}\cdot\mathbf{u}_s.
\end{equation}

Using the equation of motion derived in \S\ref{sec:3EoM}, \changes{the leading-order surface viscosity effect} on
the small-amplitude capillary wave can naturally be characterised \citep{Lopez1998}
by the \changes{non-dimensional Boussinesq number 
\begin{equation}
\mathrm{B}\equiv \mathrm{Bq}_d+\mathrm{Bq}_s = \left(\frac{\mu_d+\mu_s}{\mu}\right)k,	
\end{equation}
}where $\mathrm{Bq}_{d}=\mu_d k/\mu$ and $\mathrm{Bq}_{s}=\mu_s k/\mu$ are the Boussinesq dilatational and shear numbers, respectively, for dynamic viscosity $\mu$ and wavenumber $k$.  More explicitly, surface viscosity can be modelled to be proportional to the surfactant
concentration \citep{Ponce-Torres2017}. In the case of detergents, experimental work by \citet{Brown1953}
suggests the bi-partisan action of the special solute pairs present
in the detergent; where the primary constituent provides a large reservoir
of surface-active material while the secondary constituent, lesser
in amount, forms surface films of high viscosity. \changes{However, in the leading-order dynamics of the Boussinesq-Scriven formulation, surface viscosity is shown in \S 3 to not depend explicitly on the surfactant concentration and enters only implicitly via the surface tension coefficient.} 

\changes{The other non-dimensional numbers of the system which arise naturally in the equation of motion are the viscosity ($\epsilon$), surfactant diffusivity ($\varsigma$) and surfactant strength ($\beta$) parameters given by 
\begin{equation}
\left(\epsilon,\,\varsigma,\,\beta\right)=\frac{k}{\omega}\left(\nu k,\, D_s k,\,\frac{\alpha\Gamma_{0}}{\mu}\right),
\end{equation}
where $D_s$ denotes the coefficient of surface diffusivity, $\nu=\mu/\rho$ is the kinematic viscosity for fluid density $\rho$, $\alpha=\left|\mathrm{d}\sigma/\mathrm{d}\Gamma\right|$ is the gradient of surface tension coefficient, $\omega$ is frequency of the capillary wave and $\Gamma_0$ is the initial surfactant concentration, which is assumed to be much less than the critical micelle concentration (cmc). In this system, these parameters act as the effective Reynolds, Schmidt and Marangoni numbers, respectively.}
%Under this
%definition, $1/\mathrm{Bq}$ is merely a characteristic lengthscale
%of the measuring geometry \citep{Erni2011} and so we cannot directly
%compare $\mathrm{Bq}$ without specifying $a_{0}$. Inversely, a stability
%analysis on $\mathrm{Bq}$ must also impose geometrical conditions
%on the system.
\section{Equations of motion}\label{sec:3EoM}

The dynamics of an incompressible fluid of viscosity $\mu$ and density
$\rho$ in regions of Reynolds number $\mathrm{Re}=U\lambda/\nu\ll1$ satisfies
the Stokes equation 
\begin{eqnarray}
\rho\left(\mathbf{u}_t-\mathbf{F}\right) & = & -\boldsymbol{\nabla}p+\mu\nabla^{2}\mathbf{u}\\
\boldsymbol{\nabla}\cdot\mathbf{u} & = & 0
\end{eqnarray}
where $\mathbf{u}=(u,v)$ is the two-dimensional fluid velocity field, $p$ is
the pressure and $\mathbf{F}=-g\mathbf{j}$ is the external (gravitational)
force, with $\mathbf{j}$ denoting the upward unit vector in the $y$-direction, and $g$ is the gravitational acceleration. The small-amplitude
capillary wave is given at the free surface $F$ by the standing wave
\begin{equation}
F(x,y,t)=y-a(t)\cos kx,
\end{equation}
 where $a(t)$ is the non-linear, time-dependent wave amplitude which
satisfies the small-amplitude conditions that $a \ll\lambda\,= 2\pi/k$ and $\mathrm{d}a/\mathrm{d}t\ll v_{\mathrm{c}}=\omega/k$, 
where $\lambda$ is the wavelength and $v_{\mathrm{c}}$ is
the phase velocity. 

For vanishing Gaussian curvature in a two-dimensional space,
 the leading-order tangential and normal stress components, $\mathrm{T}_{\parallel}$
and $\mathrm{T}_{\perp}$ and the kinematic condition are given by 
\begin{eqnarray}
\mathrm{T}_{\parallel}\equiv \frac{1}{2}\mu\left(v_x+u_y\right) &  = &\boldsymbol{\nabla}_{s}\tilde{\sigma},\\
\mathrm{T}_{\perp}\equiv-p+2\mu v_y &   =&\tilde{\sigma}\boldsymbol{\nabla}_{s}\cdot\mathbf{n},\label{eq:normalstress}\\
F_t+v F_y &   =&0,
\end{eqnarray}
respectively. The leading-order normal and tangent vectors are
\begin{eqnarray}
\mathbf{n} & \simeq & (ak\sin kx,1),\\
\mathbf{t} & \simeq & (1,-ak\sin kx).
\end{eqnarray}
\changes{Similar to the small-amplitude condition, we consider a small departure from the equilibrium surface tension and let the coefficient of surface tension $\sigma$, to be defined} via a linear equation
of state 
\begin{equation}
\sigma(x,t)=\sigma_{0}-\alpha\Gamma(x,t),
\end{equation}
where $\sigma_0$ is the initial surface tension coefficient, $\Gamma(x,t)$ is the (dilute)
concentration of a surfactant solution where adsorptive-desorptive
processes are neglected.    

Using the wave-form $\Gamma(x,t)-\Gamma_{0}=\tilde{\Gamma}(t)\cos kx,$ the governing equation for surfactant concentration along a two-dimensional deforming surface \citep{Stone1990} is given by \begin{equation}
\tilde{\Gamma}_t+k^{2}D_s\tilde{\Gamma}=k\left(a_t+\nu k\Omega(0,t)*\mathcal{F}(t)\right)\label{eq:conveceqnfull}
\end{equation}
to the leading order, $\omega_{z}(x,y,t)=\Omega(y,t)\sin kx$  the \changes{$z$-component of the} vorticity, $*$ is the convolution
operator and $\mathcal{F}(t)$ is the auxiliary function 
\begin{equation}
\mathcal{F}(t)=\frac{1}{\sqrt{\pi\nu k^{2}t}}\mathrm{e}^{-\nu k^{2}t}-\mathrm{erfc}\sqrt{\nu k^{2}t}\,.
\end{equation}
The velocity and the pressure can be decomposed into inviscid and viscous
parts, i.e. $(\mathbf{u},p)=(\mathbf{u}'+\mathbf{u}'',\,p'+p''),$
where the inviscid part $(\mathbf{u}',p')$ satisfies the Euler problem
\begin{eqnarray}
\rho\left(\mathbf{u}'_t-\mathbf{F}\right) & = &-\boldsymbol{\nabla}p',\quad\label{eq:EulerStokes}\\
\left.\left(F_t+v'F_y\right)\right|_{y=0} & =& 0
\end{eqnarray}
with well-known solutions \citep{Lamb1932}
\begin{equation}
	\left(\phi,\,p'\right)=\left(\frac{1}{k}\frac{\mathrm{d}a}{\mathrm{d}t}\mathrm{e}^{ky}\cos kx,\,-\rho gy+\frac{\rho}{k}\frac{\mathrm{d}^{2}a}{\mathrm{d}t^{2}}\mathrm{e}^{ky}\cos kx\right),
\end{equation}
where $\mathbf{u}'=\boldsymbol{\nabla}\phi$. The viscous component $(\mathbf{u}'',p'')$ satisfies the Stokes
problem
\begin{eqnarray}
\rho\mathbf{u}''_t & = &\boldsymbol{\nabla}p''+\mu\nabla^{2}\mathbf{u}''\label{eq:viscousStokes}\\
0 & = & \left.v' F_y\right|_{y=0},
\end{eqnarray}
which is solved by introducing the streamfunction
$\psi$ defined by $\mathbf{u}''=(\psi_y,-\psi_x)$. Taking the curl of Eq.\,(\ref{eq:viscousStokes}) gives
the bi-harmonic equation 
\begin{equation}
(\partial_t\nabla^{2}-\nabla^4)\psi=0.
\end{equation}
Writing $\psi=\Psi(y,t)\sin kx$, the Stokes problem yields the solution
\begin{eqnarray}
2k\Psi & = & -\mathrm{e}^{-ky}\int_{-\infty}^{y}\Omega\mathrm{e}^{ky'}\mathrm{d}y'+  \mathrm{e}^{ky}\left(\int_{-\infty}^{0}\Omega\mathrm{e}^{ky'}\mathrm{d}y'+\int_{0}^{y}\Omega\mathrm{e}^{-ky'}\mathrm{d}y'\right)\\
\Omega & = &  \Omega(0,t)*\frac{y}{\sqrt{\pi\nu t^3}}\mathrm{exp}\left(-\nu k^{2}t-\frac{y^2}{4\nu t}\right)%\int_{0}^{t}\frac{y}{\sqrt{\pi\nu(t-t')^{3}}}\,\Omega(0,t') \exp\left(-\nu k^{2}(t-t')-\frac{y^{2}}{4\nu(t-t')}\right)\mathrm{d}t',
\end{eqnarray}
where we have the viscous pressure correction $p''(x,y,t)=\mu\Omega(0,t)\mathrm{e}^{ky}\cos kx$ and the boundary vorticity
\begin{equation}
	\Omega(0,t)=2\left(-\frac{\mathrm{T}_{\parallel}}{\mu}+v_x\right).
\end{equation}
Henceforth, using non-dimensional variables $\tau=\omega t,$ $\epsilon=\nu k^{2}/\omega$
and $\tilde{\Omega}=\Omega(0,t)/\omega$, the boundary vorticity becomes the integral equation 
\begin{equation}
\tilde{\Omega}(0,\tau) = \mathrm{f}(\tau)+2\epsilon\,\mathrm{B}\,\tilde{\Omega}(0,\tau')*\mathcal{F}(\tau)\label{eq:VORT-EQN}
\end{equation}
where we have 
\begin{equation}
	\mathrm{f}(\tau)=-2[\beta\tilde{\Gamma}(\tau)+\delta(1+\mathrm{B})\dot{A}],
\end{equation}
for $A=a/a_0$ is the dimensionless amplitude, $\delta=a_{0}k$ and $\dot{}=\mathrm{d}/\mathrm{d}\tau$ denotes the non-dimensional temporal
derivative. Substituting the pressure and the velocity into Eq.\,(\ref{eq:normalstress})
and Eq.\,(\ref{eq:conveceqnfull}) gives the simultaneous
equation 
\begin{eqnarray}
\ddot{A}+2\epsilon\dot{A}+A & = & \epsilon\tilde{\Omega}(0,\tau)-2\epsilon^{2}\tilde{\Omega}(0,\tau)*\mathcal{F}(\tau),\label{eq:AMPLITUDE-EQN}\\
\dot{\tilde{\Gamma}}+\varsigma\tilde{\Gamma} & = & \delta\dot{A}+\epsilon\tilde{\Omega}(0,\tau)*\mathcal{F}(\tau).\label{eq:SURFACTANT-EQN}
\end{eqnarray}
Equations (\ref{eq:AMPLITUDE-EQN}) and (\ref{eq:SURFACTANT-EQN}) provides us with a dynamic equation system for the amplitude and the surfactant concentration, the solution of which we outline in the next section. 

\section{Solution of the simultaneous integro-differential equation}\label{sec:4solution}

Let $F(s)=\mathcal{L}[A](s)$, $G(s)=\mathcal{L}[\tilde{\Gamma}](s)$ \changes{and $\hat{\Pi}(s)=sF(s)-A_0$}
be the Laplace transforms of $A(\tau),\,\tilde{\Gamma}(\tau)$ and $\dot{A}(\tau)$,
define the polynomial expressions $\Theta_{\epsilon}^{(i)}\equiv\Theta_{\epsilon}^{(i)}(s+\epsilon)$ \changes{
for $1\leqslant i\leqslant6$ as
\begin{eqnarray}
\Theta_{\epsilon}^{(1)} & = &2\left[(s+\epsilon)^{1/2}-\epsilon^{1/2}\right],\\
\Theta_{\epsilon}^{(2)} & = &s\epsilon^{1/2}+\mathrm{B}\epsilon\Theta^{(1)},\\
\Theta_{\epsilon}^{(3)} & = &(s+\varsigma)\Theta^{(2)}+\beta\epsilon\Theta^{(1)},\\
\Theta_{\epsilon}^{(4)} & = &(s^{2}+2\epsilon s+1+2\mathrm{B}'\epsilon s)\Theta^{(2)}-2\epsilon^{2}s\mathrm{B}'^{2}\Theta^{(1)},\\
\Theta_{\epsilon}^{(5)} & = &2\epsilon s\beta\left(\mathrm{B}'\epsilon\Theta^{(1)}-\Theta^{(2)}\right),\\
\Theta_{\epsilon}^{(6)} & = &\delta\Theta^{(2)}-\mathrm{B}'\epsilon\Theta^{(1)},
\end{eqnarray}
where $\mathrm{B}'=1+\mathrm{B}$.} The rational function $\hat{\Pi}_{\epsilon}=\hat{\Pi}_{\epsilon}(s+\epsilon)=\mathrm{P}(s^{1/2})/\mathrm{Q}(s^{1/2})$
is decomposed into its partial fraction  
\begin{eqnarray}
\hat{\Pi}_{\epsilon}(s+\epsilon)  & \equiv&\sum_{i=1}^{10}\frac{c_{i}}{s^{1/2}+z_{i}}\label{eq:partialdecomp} \\
& = &  \frac{(U_{0}s-A_{0})\Theta^{(2)}\Theta^{(3)}+\tilde{\Gamma}_{0}\Theta^{(2)}\Theta^{(5)}}{\Theta^{(4)}\Theta^{(3)}-\tilde{\Gamma}_{0}\Theta^{(5)}\Theta^{(6)}},\label{eq:PI}
\end{eqnarray}
where $-z_{i}$ are the roots of the polynomial $\mathrm{Q}(s^{1/2})$. 
\changes{
In the absence of the Marangoni effect, the surface viscosity case is given by 
\begin{equation}
\hat{\Pi}_{\epsilon}(s+\epsilon)=\frac{(U_{0}s-A_{0})\Theta^{(2)}}{(s^{2}+2\epsilon s+1+2\mathrm{B}'\epsilon s)\Theta^{(2)}-2\epsilon^{2}s\mathrm{B}'^{2}\Theta^{(1)}}.	
\end{equation}
}
By comparison with Lagrange polynomial \changes{interpolation, we have 
\begin{eqnarray}
\frac{\mathrm{P}(s^{1/2})}{\mathrm{Q}(s^{1/2})}&\equiv & \mathrm{P}(s^{1/2})\prod_{i=1}^{k}\frac{1}{s^{1/2}+z_{i}}\\
&=&\sum_{i=1}^{k}\frac{\mathrm{P}(-z_{i})}{\sigma_{i}^{(k)}(-z_{i})}\frac{1}{s^{1/2}+z_{i}}, \label{eq:lagrangeinterp}
\end{eqnarray}}
where $\sigma_{i}^{(n)}$ is the $n$-th order cyclic polynomial given
by 
\begin{equation}
\sigma_{j}^{(n)}=\prod_{i=1}^{n-1}\left(z_{j+i\,\mathrm{mod}(n)}-z_{j}\right).
\end{equation}
It follows that by comparing Eq.\,(\ref{eq:partialdecomp}) and Eq.\,(\ref{eq:lagrangeinterp}), the coefficients $c_{i}$
are 
$c_{i}=\mathrm{P}(-z_{i})/\sigma_{i}^{(10)}(-z_{i})$. \changes{Let \begin{equation}
\mathrm{Z}(n,j)=\sum_{i=1}^{n}\frac{\mathrm{P}(-z_{i})}{\sigma_{i}^{(n)}}(-z_{i})^{j};
\end{equation}
it follows (see Appendix A) that the condition 
\begin{equation}
\deg\mathrm{Q}-\deg\mathrm{P}=2	
\end{equation}
implies $\mathrm{Z}(n,0)=0$, where $\deg X$ is the degree of the polynomial $X$. }Taking the inverse Laplace transform of Eq. (\ref{eq:PI}) gives \changes{
\begin{equation}
\Pi_{\epsilon}(\tau)  =\frac{\mathrm{Z}(10,0)}{\sqrt{\pi t}}-\sum_{i=1}^{10}\frac{\mathrm{P}(-z_{i})}{\sigma_{i}^{(10)}}z_{i}\mathrm{e}^{z_{i}^{2}\tau}\mathrm{erfc}(z_{i}\tau^{1/2}).
\end{equation}}
Finally, the non-dimensional amplitude is given by
\begin{equation}
A(\tau)=1+\sum_{i=1}^{10}\frac{z_{i}}{\sigma_{i}^{(10)}}\mathrm{P}(-z_{i})\varphi(z_{i},\tau;\epsilon),\label{eq:AMP-SOL}
\end{equation}
where $\varphi=\varphi(z_{i},\tau;\epsilon)$ satisfies 
\begin{equation}
	\varphi(z_{i},\tau;\epsilon)=\frac{1}{z_i^2-\epsilon}\left(\mathrm{e}^{(z_{i}^{2}-\epsilon)\tau}\mathrm{erfc}(z_{i}\tau^{1/2})+\frac{z_{i}}{\epsilon^{1/2}}\mathrm{erf}[(\epsilon\,\tau)^{1/2}]-1\right).
\end{equation}

\section{Surface viscosity effects on the wave amplitude}\label{sec:5wavedisperse}

\begin{figure}
\subfloat[$\lambda=0.95\lambda_c^{(0)}$]{\includegraphics[width=3.25cm]{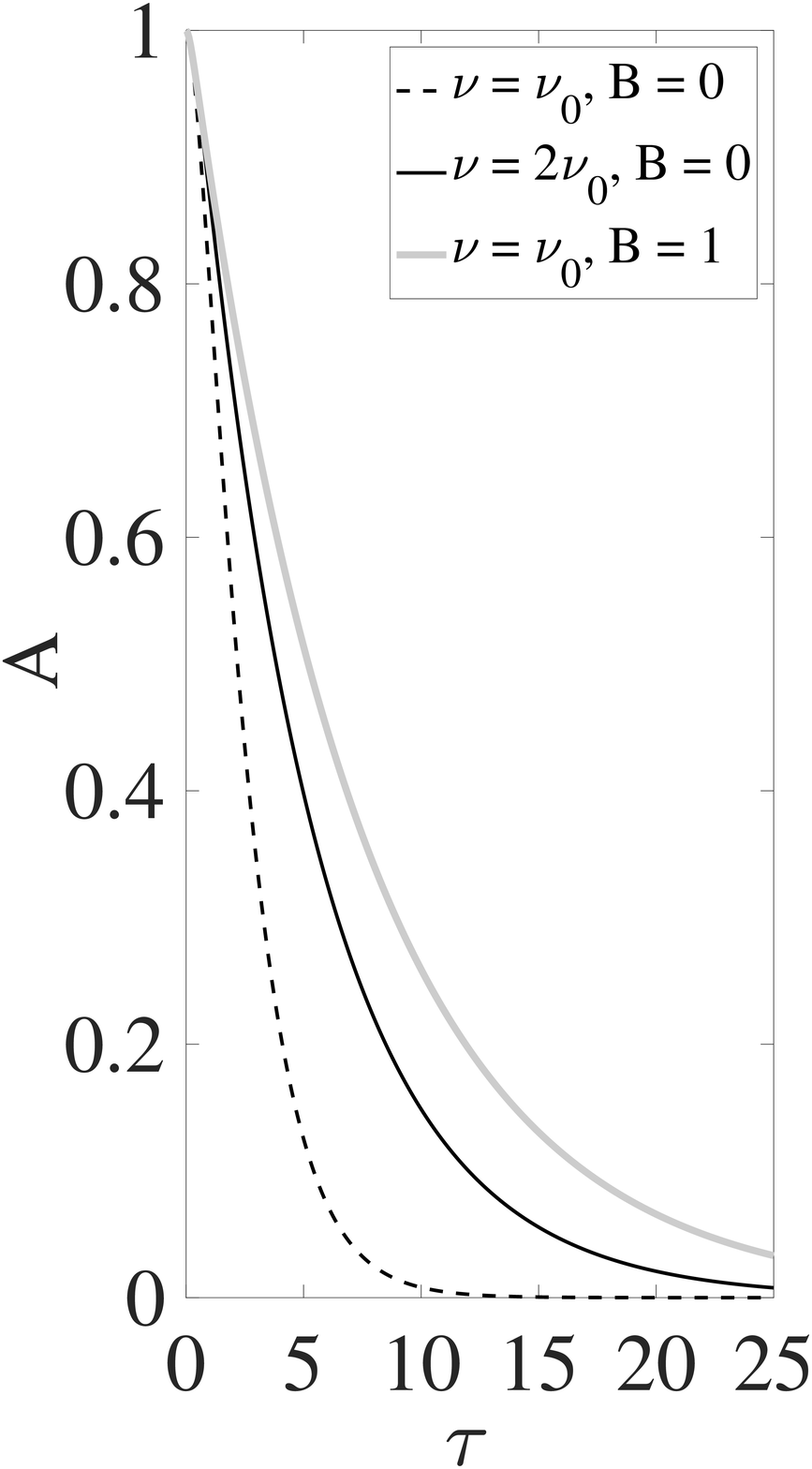}}
\subfloat[$\lambda=9.5\lambda_c^{(0)}$]{\includegraphics[width=3.25cm]{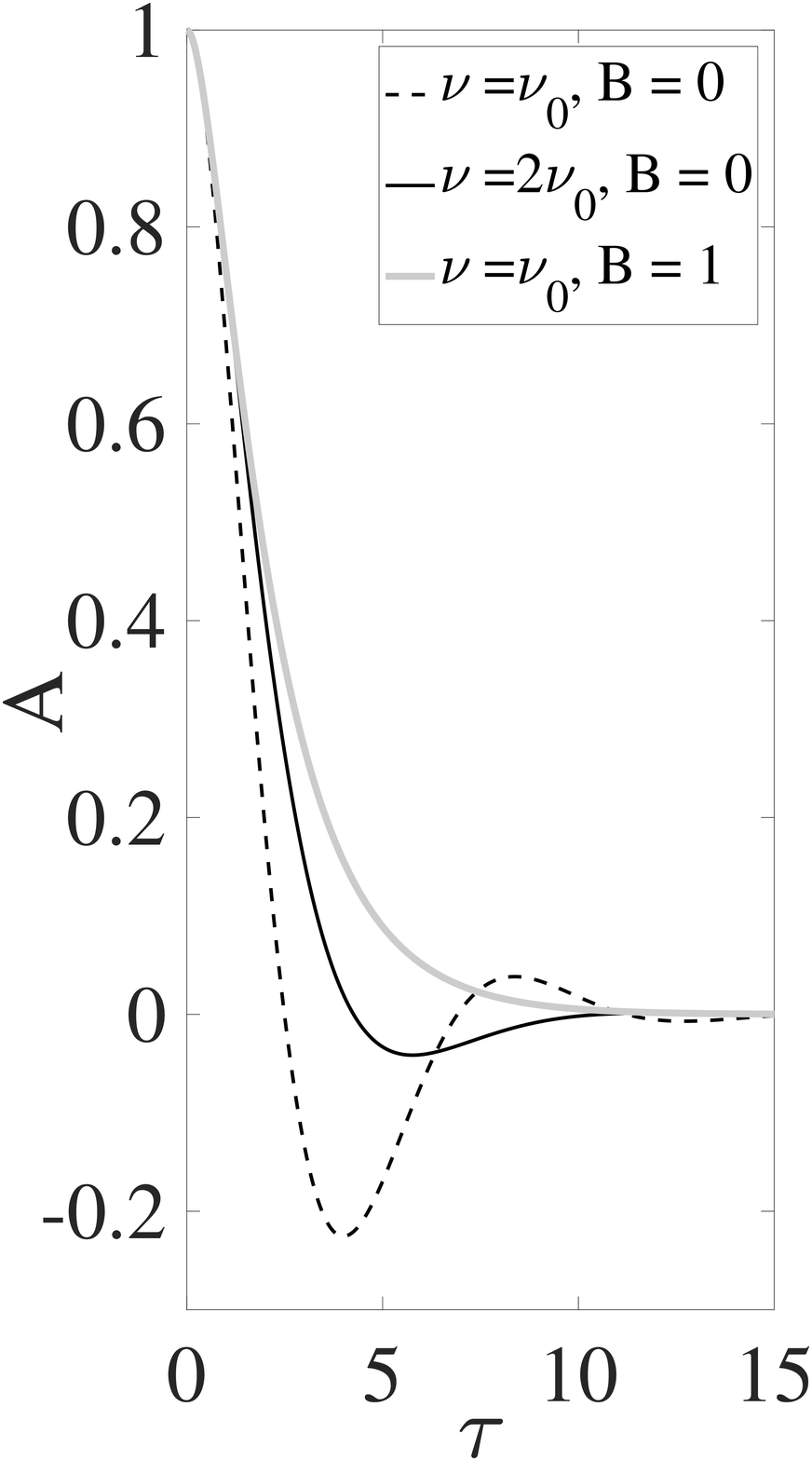}}
\subfloat[$\lambda=95\lambda_c^{(0)}$]{\includegraphics[width=3.25cm]{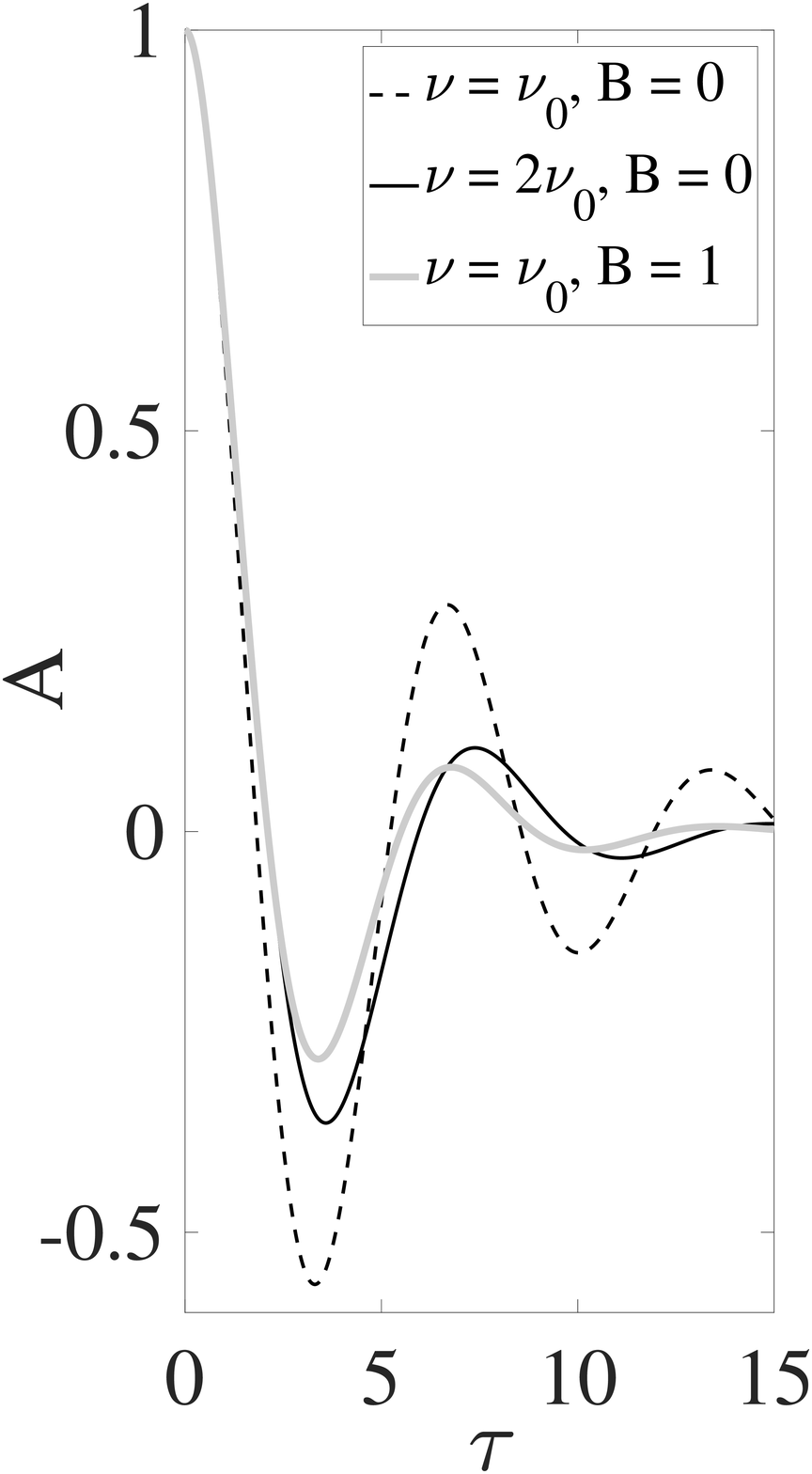}}
\subfloat[$\lambda=950\lambda_c^{(0)}$]{\includegraphics[width=3.25cm]{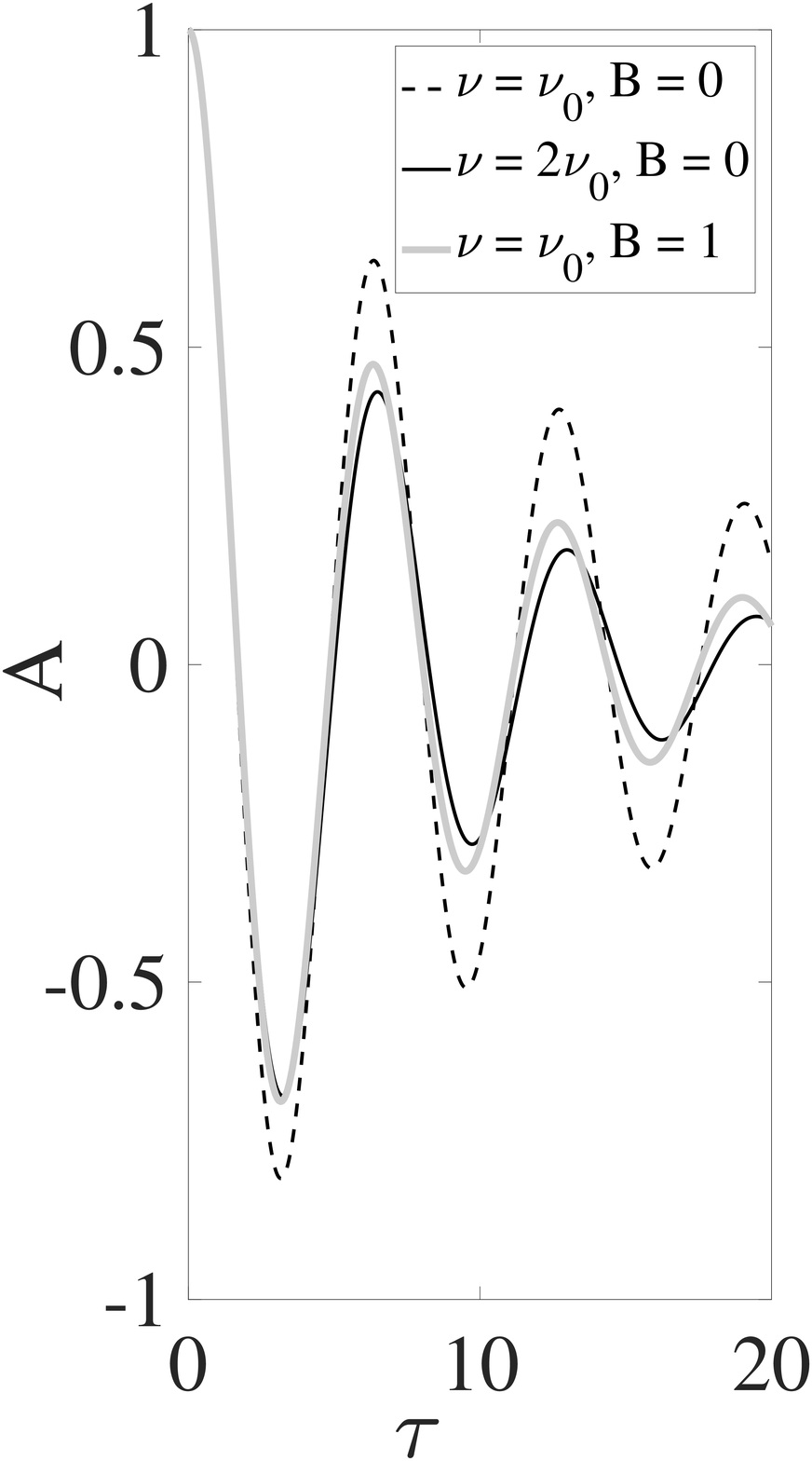}}

\subfloat[$\lambda=0.95\lambda_c^{(0)}$]{\includegraphics[width=3.25cm]{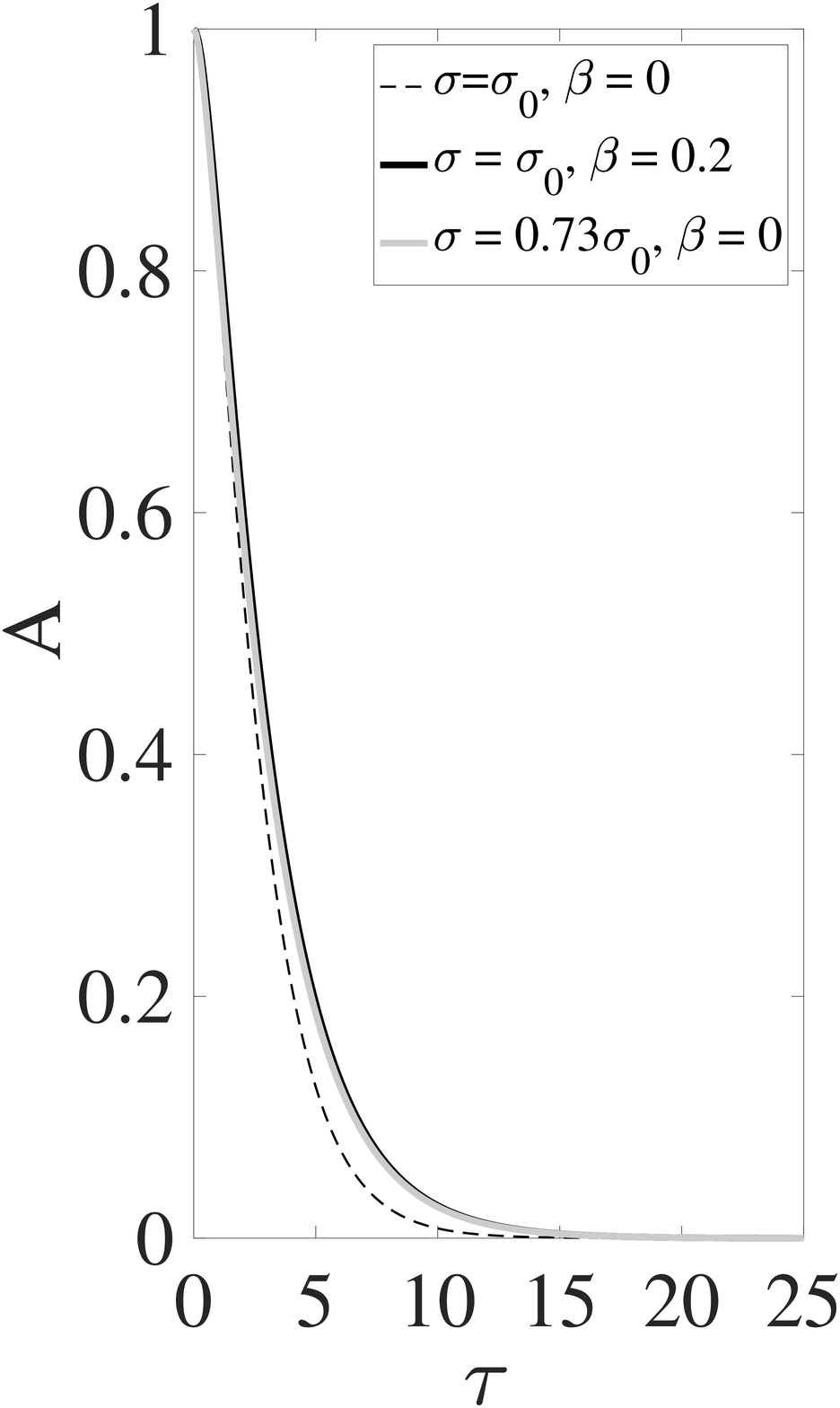}}
\subfloat[$\lambda=9.5\lambda_c^{(0)}$]{\includegraphics[width=3.25cm]{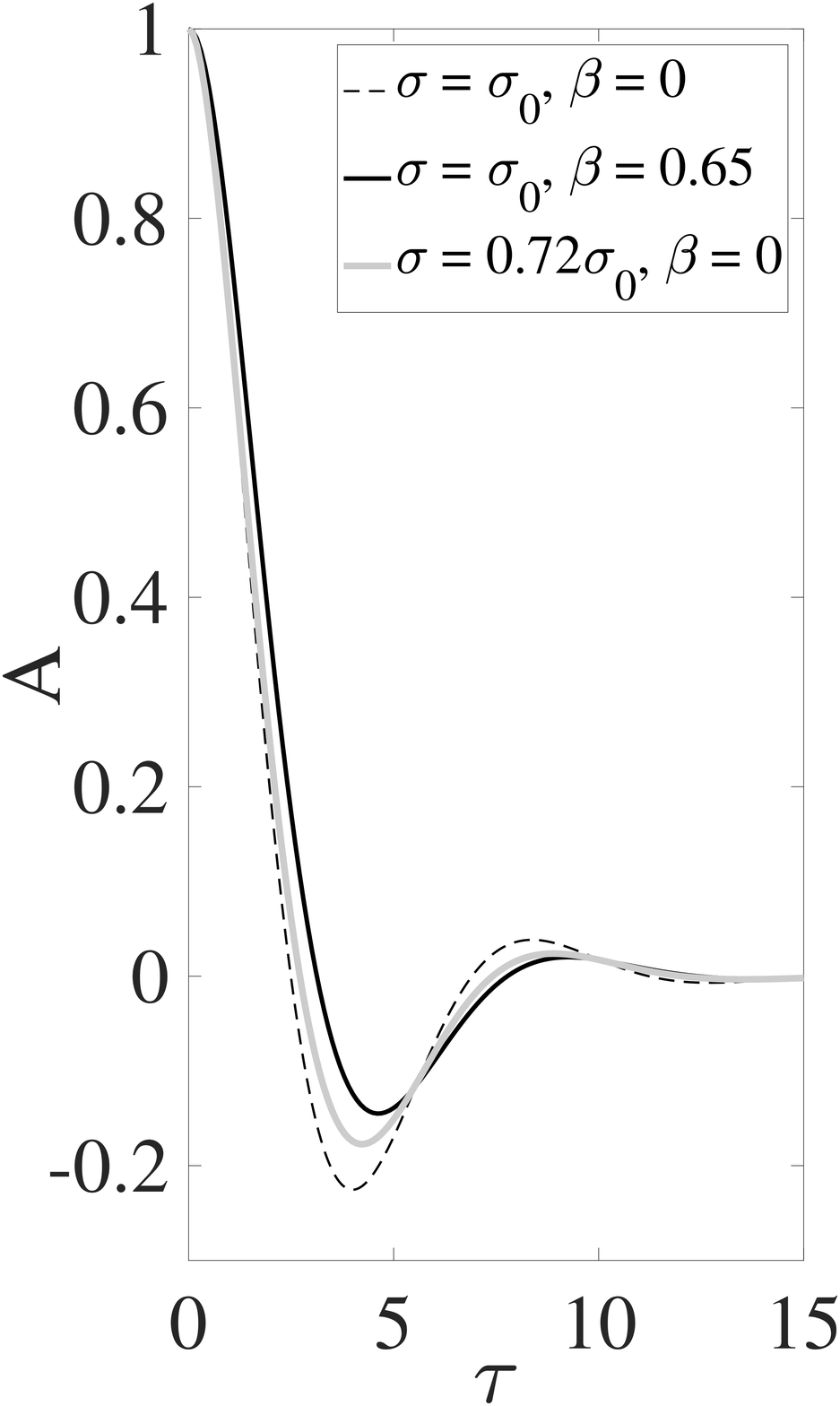}}
\subfloat[$\lambda=95\lambda_c^{(0)}$]{\includegraphics[width=3.25cm]{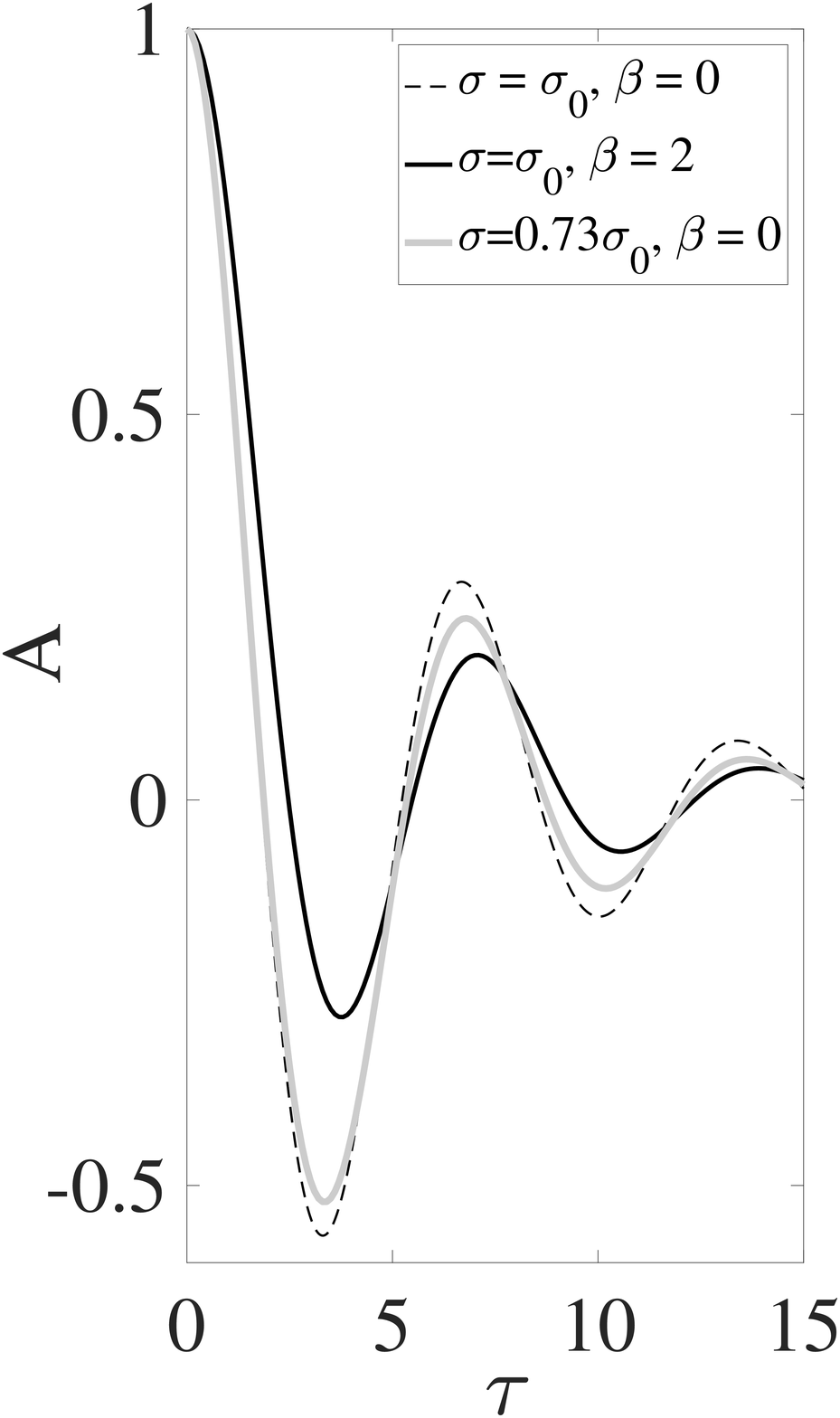}}
\subfloat[$\lambda=950\lambda_c^{(0)}$]{\includegraphics[width=3.25cm]{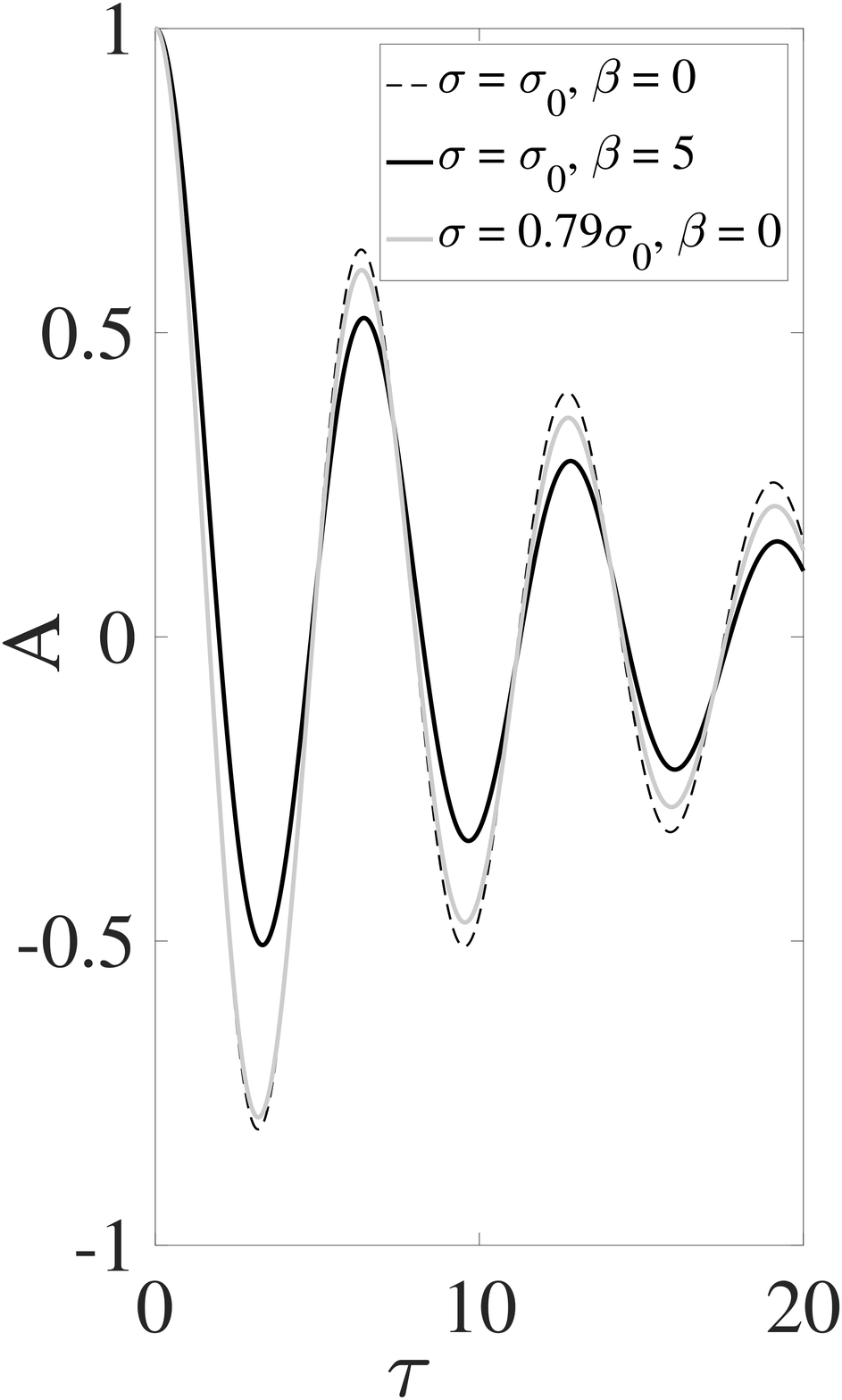}}
\caption{Wave amplitude as a function of dimensionless time $\tau$
for water at approximate room temperature and pressure for \changes{$\mathrm{Sc}=10^4$}, comparing the influence of the surface viscosity and Marangoni effects for various values of $\mathrm{B}$ (with $\beta=0$) and $\beta$ (with $\mathrm{B}=0$). Here $\nu_0=8.9\times10^{-7}\mathrm{m^2s^{-1}}$ and $\sigma_0=0.072\mathrm{Nm^{-1}}$ denotes the baseline kinematic viscosity and surface tension, respectively. The modified surface tension in subfigures e-f represent a clean system (i.e. $\beta=\mathrm{B}=0$) with $\sigma=\sigma_0-\alpha\Gamma_0$.}
\end{figure}

\begin{figure}
\subfloat[$\lambda=0.95\lambda_c^{(0)}$]{\includegraphics[width=3.25cm]{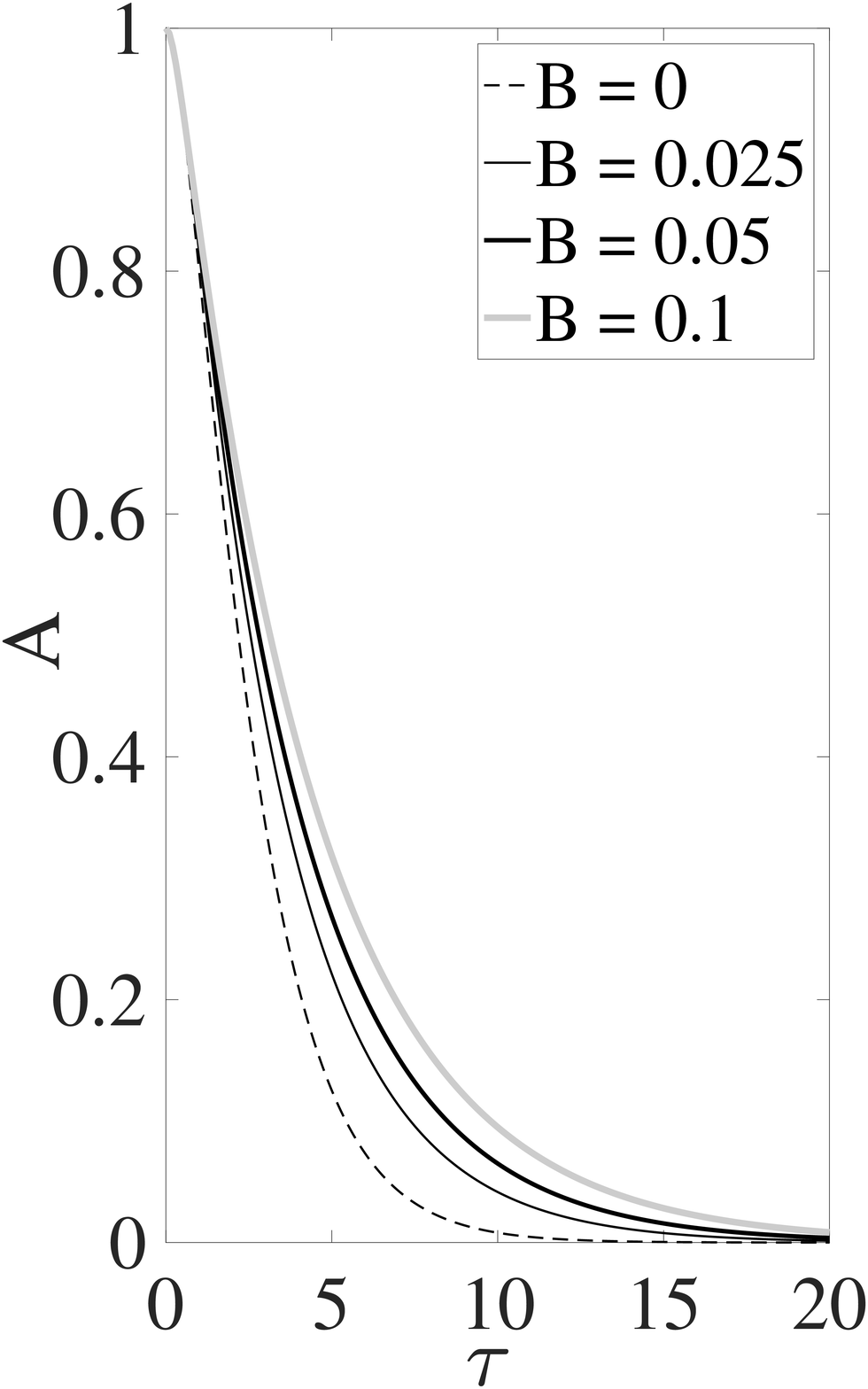}}
\subfloat[$\lambda=0.95\lambda_c^{(0)}$]{\includegraphics[width=3.25cm]{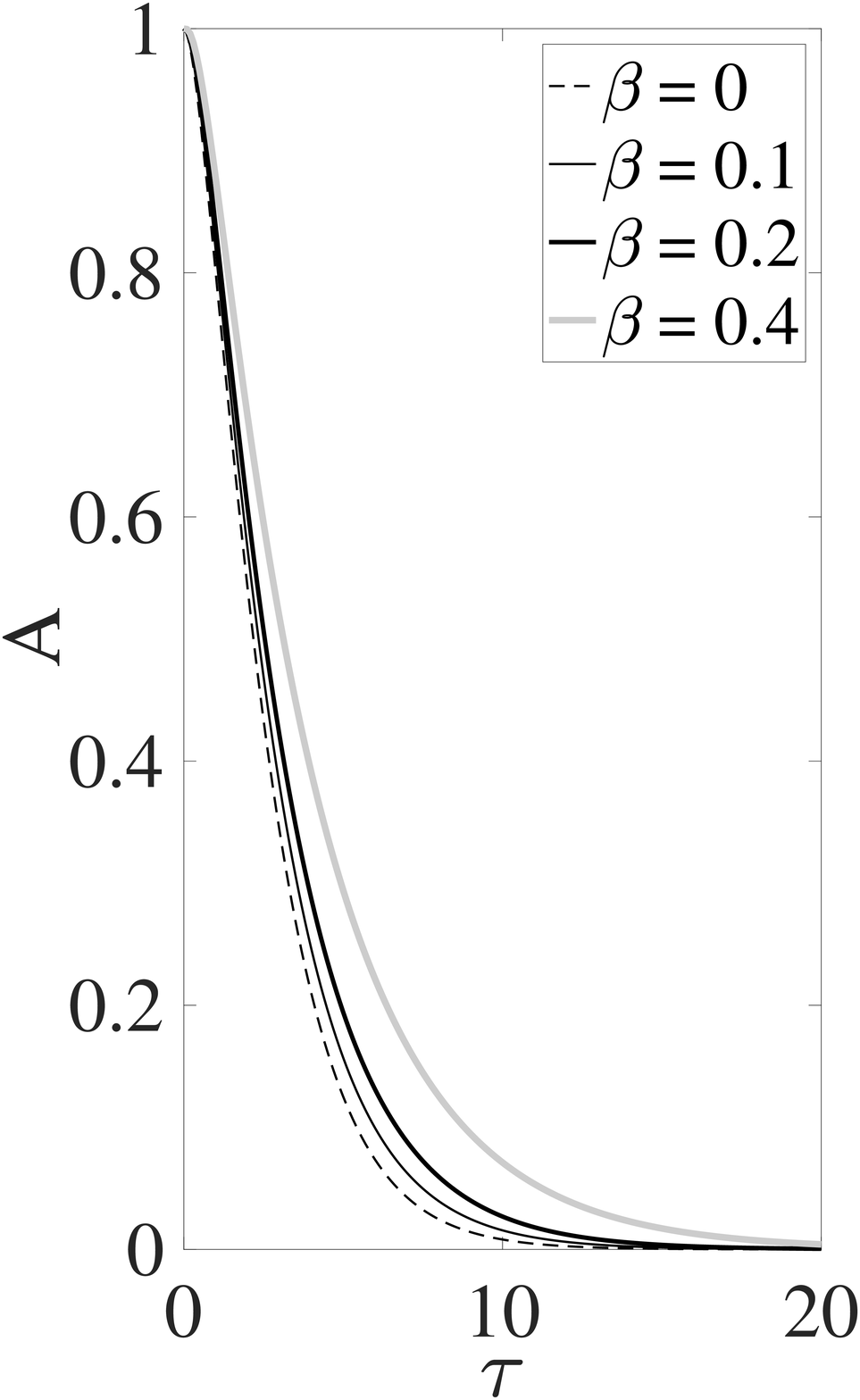}}
\subfloat[$\lambda=9.5\lambda_c^{(0)}$]{\includegraphics[width=3.25cm]{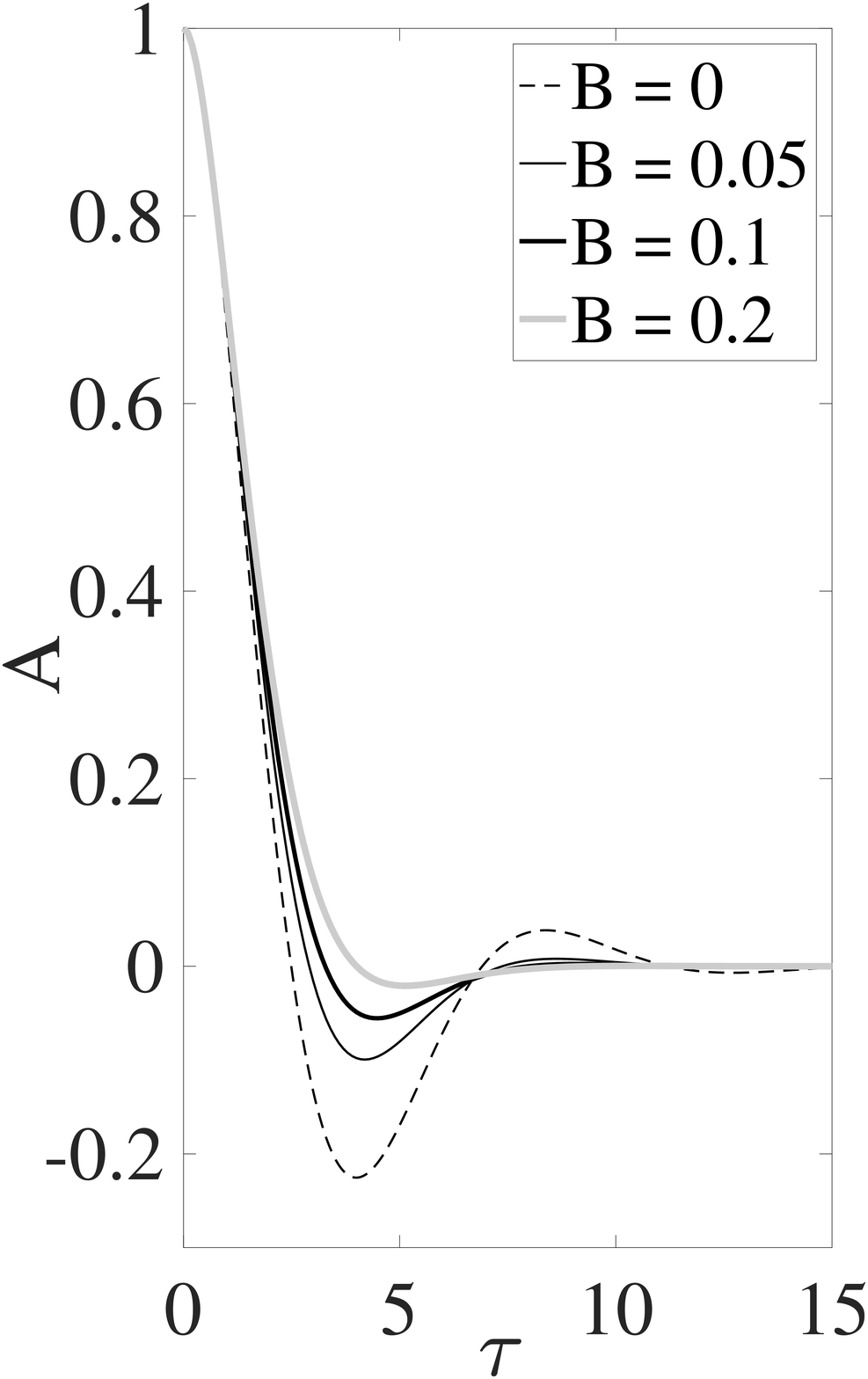}}
\subfloat[$\lambda=9.5\lambda_c^{(0)}$]{\includegraphics[width=3.25cm]{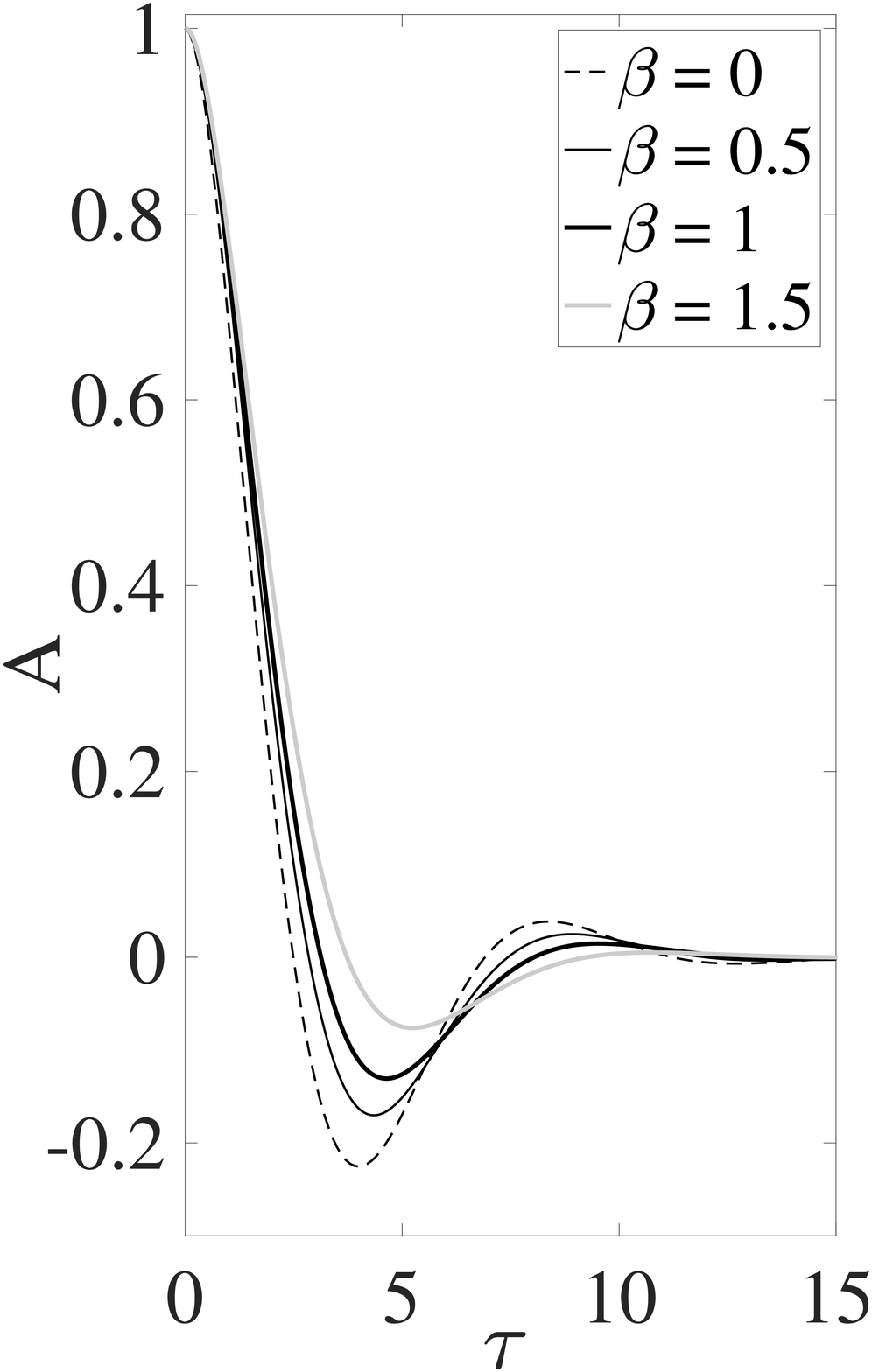}}

\subfloat[$\lambda=95\lambda_c^{(0)}$]{\includegraphics[width=6.5cm]{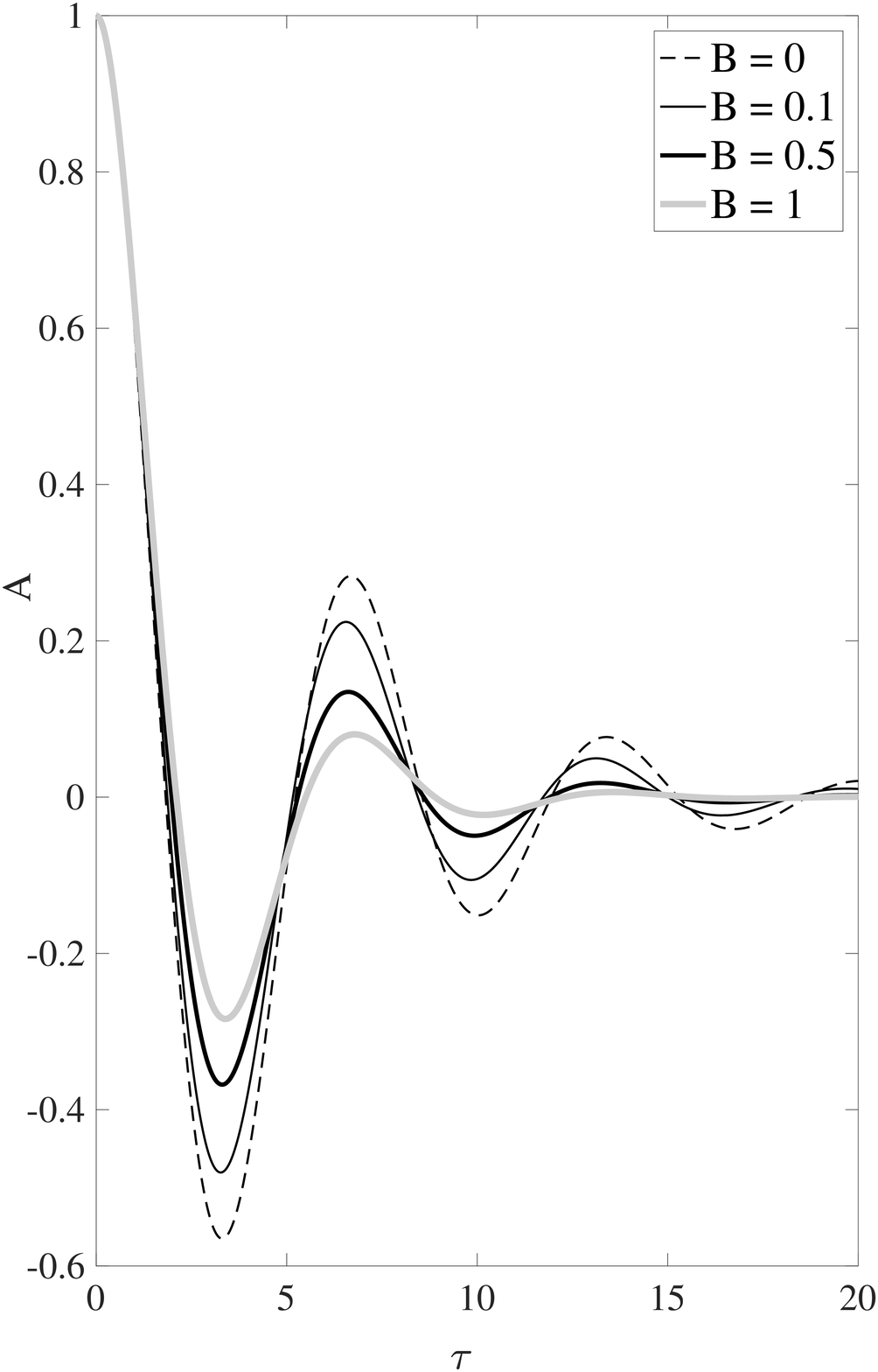}}
\subfloat[$\lambda=95\lambda_c^{(0)}$]{\includegraphics[width=6.5cm]{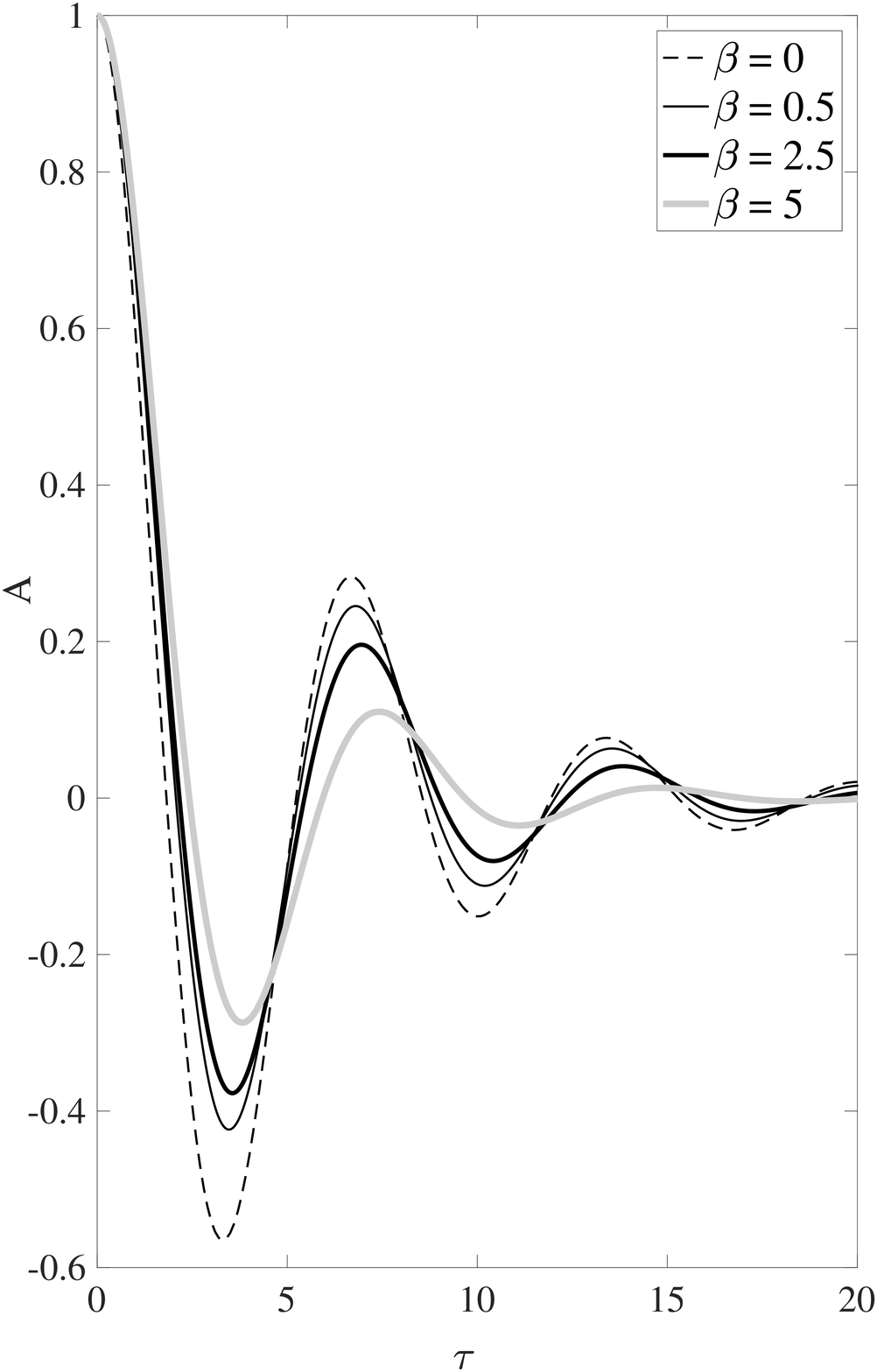}}

\caption{Wave amplitude as a function of dimensionless time $\tau$ 
for water at approximate room temperature and pressure for \changes{$\mathrm{Sc}=10^4$}, showing the influence of the surface viscosity and Marangoni effects for various values of $\mathrm{B}$ (with $\beta=0$) and $\beta$ (with $\mathrm{B}=0$) for different wavelength $\lambda$. }
\end{figure}

As shown in \S4, the leading-order surface viscosity effects on
the dynamics of small-amplitude capillary waves are characterised
by the parameter $\mathrm{B}$ and similarly, the Marangoni effect can
be reduced to the non-dimensional variables $\varsigma$ and $\beta$ given in \S 2. In
what follows, we use water under room temperature and pressure (rtp),
i.e. at 25$^\circ$C, with density $\rho=10^{3}\mathrm{kgm}^{-3}$,
surface tension $\sigma=7.2\times10^{-2}\mathrm{Nm}^{-1}$ and viscosity
\changes{$\mu=8.9\times10^{-4}\mathrm{Pa\,s}$, as a test system with surfactant Schmidt number $\mathrm{Sc}=\nu/D_s=10^4$}. We define $\lambda_c^{(0)}$ to be the wavelength for which the capillary wave undergoes critical damping, henceforth known as the critical wavelength. The superscript denotes the clean case which we understand as a system without the addition of surface-active material, i.e. $\beta=\mathrm{B}=0$. We will look at this critical wavelength in more detail in \S7, but here we note a harmonic oscillator approximation of $\lambda_c^{(0)}$   \citep{Denner2016b} whereby 
\begin{equation}
\lambda_{\mathrm{c}}^{(0)}=2^{1/3}\pi l_{\mathrm{vc}}/\Theta,\label{eq:HARMONIC-Crit}
\end{equation}
for  $\Theta=1.0625$ and the viscocapillary lengthscale 
\begin{equation}
	l_{\mathrm{vc}}=\frac{\mu^{2}}{\rho\sigma}.
\end{equation}

\changes{In figure 1, we compare the effect of surface
viscosity with that of the equivalent bulk viscosity $\nu$, and the
Marangoni effect with that of simple reductions in the surface tension coefficient $\sigma=\sigma_0-\alpha\Gamma_0$. Here we use the phrase equivalent to denote the quantity of bulk viscosity and simple reductions in surface tension which results in an identical effect on the overall wave amplitude due to surface viscosity and the Marangoni effect, respectively, under the limit of either a large or small wavelength. For wavelengths
near critical damping, in figure 1(a), increasing $\mathrm{B}$ exhibits a relatively
large difference to equivalent increases in $\nu$, as compared to the case for $\lambda=950\lambda_c^{(0)}$ in figure 1(d). This difference
decreases as we increase the wavelength into less damped regions
as shown in figure 1(b) and 1(c). In contrast, the picture is reversed for the Marangoni
effect, where equivalent changes in the surface tension coefficient are almost identical
to increases in surfactant concentration (through $\beta$) near critical damping in figure 1(e) and
their difference increases with larger wavelength from figure 1(f) to figure 1(h); where the dynamic Marangoni effect vastly overshadows the static changes in the surface tension coefficient. Henceforth, we can approximate the Marangoni effect on the wave amplitude near the critical wavelength with the equivalent reduction in $\sigma$. 

%Intuitively
%speaking, we note that the fluid flow
%near critical damping is almost quasi-static, so the surfactant is not being transported rapidly
%enough to induce significant gradient near critical damping and thus for $\lambda\sim \lambda_c^{(0)}$ the true Marangoni effect on the wave amplitude does not deviate hugely from the equivalent simple reductions in surface tension. For larger wavelengths, the flow is not quasi-static and thus the surfactant is transported by the significant fluid advection creating a large gradient in the surface tension coefficient and which the overall effect on wave amplitude cannot be described by a simple reduction of the surface tension gradient alone. On the other hand, for the surface viscosity effect and its equivalency to increases in the fluid bulk viscosity, we consider that from the definition of $\mathrm{B}=(\mu_d+\mu_s)k/\mu$, by maintaining $\mathrm{B}=1$ for decreasing $k$ in figures 1(a-d), we effectively increases the proportion of $(\mu_d+\mu_s)/\mu$ for larger wavelengths, leading to near parity with the doubling of bulk viscosity for $\lambda\gg \lambda_c^{(0)}$.

In figure 2, we see in more detail the mechanisms of surface viscosity and the Marangoni effect at altering wave amplitudes for different wavelengths. Both effects admit relatively self-similar solutions and that increasing either the surface viscosity or the surfactant concentration increases damping as expected. One of the explicit differences is that the Marangoni effect reduces surface tension and, thus, lowers the frequency $\omega$, while surface viscosity leaves $\omega$ unchanged. This is due to surface viscosity being dependent on the surfactant concentration only to the linear order, and does not feature in the leading-order nonlinear amplitude equation in Eq.(\ref{eq:AMPLITUDE-EQN}) explicitly, as noted previously. The consequence on the amplitude of this $\omega$-lowering is a horizontal drift of the waveform for systems with the Marangoni effect, as evident from figures 2(d) and 2(f), as opposed to relatively centred waveforms in the cases of surface viscosity in figures 2(c) and 2(e). 

We also observe that the surface viscosity effect weakens for  large wavelengths $\lambda \gg \lambda_c$ but is very potent for small wavelengths $\lambda \sim \lambda_c$. A particular consequence of this potency  is its ability to alter the onset behaviour in interfacial phenomena, a number of which occurs near the region of the critical wavelength. The stochastic  nature of many of the interfacial instabilities \citep{Aarts2004} are often kickstarted by the small-amplitude local disturbances on the interface. Hence a small change in surface material, and thus the wave damping, can have a significant effect in starting or delaying the initialisation process of more complex phenomena. Furthermore, it would be of interest to obtain the modifications to the critical wavelength upon the addition of a small amount surface active material. However, we need to consider the definition of the critical wavelength for a higher-order system as it is not as readily defined as in a second-order system. }

\section{Capillary wave dispersion and the critical wavelength}
To quantify the changes due to the presence of the Marangoni and surface viscosity effects to the critical wavelength, we must first obtain the critical wavelength in the clean case. \changes{Following \citet{Lamb1932}, the general dispersion relation for an interface with both the Marangoni effect and surface viscosity can be found (derivation in Appendix B)
 to take the form \begin{equation}
W_{0}(Z;\epsilon)\left(1+Z+\mathrm{B}+\frac{\beta}{\epsilon^{2}(Z^{2}-1)}\right)+\left(\mathrm{B}+\frac{\beta}{\epsilon^{2}(Z^{2}-1)}\right)(Z-1)^{3}=0\label{eq:generaldisp}
\end{equation}
where 
\begin{equation}
\frac{\mathrm{i}\omega}{\epsilon}=Z^{2}-1
\end{equation}
and 
\begin{equation}
W_{0}(Z;\epsilon)=Z^{4}+2Z^{2}-4Z+1+\frac{1}{\epsilon^{2}}.
\end{equation}}Specialising to the clean case, Eq.\,(\ref{eq:generaldisp}) reduces to
%\begin{equation}
%\omega_{0}^{2}+(\mathrm{i}\omega+2\nu k^{2})^{2}-4\nu^{2}k^{4}\left(1+\frac{\mathrm{i}\omega}{\nu k^{2}}\right)^{1/2}=0.\label{eq:dispersion-relation}
%\end{equation}
\begin{equation}
(\mathrm{i}\omega'+2\epsilon)^{2}-4\epsilon^2\left(1+\frac{\mathrm{i}\omega'}{\epsilon}\right)^{1/2} + 1 =0,\label{eq:dispersion-relation}
\end{equation}
as derived from linearised hydrodynamics \citep{Levich1962,Lamb1932}. The dispersion relation admits solution of the form 
\begin{equation}
\omega'=2\mathrm{i}\epsilon-\frac{1}{2}h-\left(1-\frac{1}{4}h^{2}-\frac{8\mathrm{i}\epsilon^{3}}{h}\right)^{1/2},\label{eq:omega-root}
\end{equation}
where
$h^2  = \frac{1}{3} -  J_{+}^{1/3} - J_{-}^{1/3}$ for 
\begin{equation}
J_{\pm}= \frac{1}{27}-\frac{2}{3}\epsilon^{4}+2\epsilon^{6} \pm \frac{2}{3\sqrt{3}}\epsilon^{3}\sqrt{\mathrm{f}(\epsilon)},
\end{equation}
where the polynomial $\mathrm{f}(\epsilon)$ is given by 
\begin{equation}
\mathrm{f}(\epsilon)=11\epsilon^{6}-18\epsilon^{4}-\epsilon^{2}-1.\label{eq:transitionpoly}
\end{equation}
For $\epsilon\ll\epsilon^{\star},$ the wave
frequency can be written as 
\begin{equation}
\omega'  \sim\frac{1}{2}h^2-h+2\mathrm{i}\epsilon\left(1+\frac{\epsilon^{2}}{h}\right),
\end{equation}
and the damping coefficient can be extracted as $\mathrm{Im}(\omega')\sim2\epsilon,\label{eq:damping-low-viscosity}$ where $\epsilon^{\star}\in\mathbb{R}^{+}$ is the transition value defined by the (largest positive) root of the
polynomial $\mathrm{f},$ i.e. 
\begin{equation}
\epsilon^{\star}=\sup_{\mathbb{R}^{+}}\left\{ \epsilon:\mathrm{f}(\epsilon)=0\right\}. 
\end{equation}
Solving Eq.\,(\ref{eq:transitionpoly}) exactly gives 
\begin{equation}
\epsilon^\star = \left(\frac{6}{11}+\frac{1}{33}\left(\frac{3}{2}(5571-341\sqrt{93}\right)^{1/3}+\frac{1}{33}\left(\frac{3}{2}(5571+341\sqrt{93}\right)^{1/3}\right)^{1/2}	
\end{equation}
with the numerical value $\epsilon^{\star}\simeq1.3115.$ \changes{Reintroducing dimensional variables, define $P,Q$ and the variable $K$ by} % $P=-3(k^\star -K)^2$ and $Q=\mathrm{g}\epsilon^{\star2}/\nu^{2} + 2\sigma^{3}\epsilon^{\star6}/27\rho^{3}\nu^{6}$.
\changes{\begin{eqnarray}
K & = & k^{\star}-\frac{\sigma\epsilon^{\star2}}{3\nu^{2}\rho}\\
P & = & -\frac{\sigma^{2}\epsilon^{\star4}}{3\rho^{2}\nu^{4}}\\
Q & = & -\left(\frac{\mathrm{g}\epsilon^{\star2}}{\nu^{2}}+\frac{2\sigma^{3}\sigma^{\star6}}{27\rho^{3}\nu^{6}}\right).
\end{eqnarray}}It follows from the definition of $\epsilon$ that the critical wavenumber $k^{\star}$ satisfies the
cubic equation 
\begin{equation}
K^{3}+PK+Q=0.
\end{equation}
We require the real solution given by 
\begin{equation}
K=\left\{ -\frac{1}{2}Q+\Delta^{1/2}\right\} ^{1/3}+\left\{ -\frac{1}{2}Q-\Delta^{1/2}\right\} ^{1/3},\label{eq:k-star-solution}
\end{equation}
where 
\begin{eqnarray}
\Delta^{1/2} & \equiv & \left(\frac{Q^{2}}{4}+\frac{P^{3}}{27}\right)^{1/2}\\
 & = & \frac{\mathrm{g}\epsilon^{\star2}}{2\nu^{2}}\left(1+\frac{4\sigma^{3}\epsilon^{\star4}}{27\rho^{3}\nu^{4}\mathrm{g}}\right)^{1/2}.
\end{eqnarray}
%\begin{equation}
%K =\left\{-\frac{Q}{2}+\frac{\mathrm{g}\epsilon^{\star2}}{2\nu^{2}}\left(1+\frac{4\sigma^{3}\epsilon^{\star4}}{27\rho^{3}\nu^{4}\mathrm{g}}\right)^{1/2}\right\} ^{1/3}+\left\{-\frac{Q}{2}-\frac{\mathrm{g}\epsilon^{\star2}}{2\nu^{2}}\left(1+\frac{4\sigma^{3}\epsilon^{\star4}}{27\rho^{3}\nu^{4}\mathrm{g}}\right)^{1/2}\right\} ^{1/3}.\label{eq:k-star-solution}
%\end{equation}
By inspecting Eq.\,(\ref{eq:k-star-solution}), the critical wavenumber
under the limit $k\ll (\rho \mathrm{g}/\sigma)^{1/2}$ is $k^{\star}\sim\left(\mathrm{g}\epsilon^{\star2}/\nu^{2}\right)^{1/3},$ corresponding to the gravity-dominated regime with $\omega_{0}\sim(\mathrm{g}k)^{1/2}$. For $k\geqslant(\rho\mathrm{g}/\sigma)^{1/2}$,
the critical wavenumber reduces to
\begin{equation}
k^{\star}\sim\epsilon^{\star2}\frac{\sigma\rho}{\mu^{2}}=\frac{\epsilon^{\star2}}{l_{\mathrm{vc}}},\label{eq:CRIT-ANALYTIC}
\end{equation}
corresponding to the capillary-dominated regime with $\omega_{0}\sim(\sigma k^{3}/\rho)^{1/2}$. 

For $\epsilon\gg\epsilon^{\star}$,
we note that $\mathrm{Re}(\omega')=0$ and the system is in an overall
overdamped regime. The damping ratio is given by expanding $\omega_{-}'$
(since $\omega_{+}'\gg\omega_{-}'$ and would thus rapidly damp the
motion) in Eq.\,(\ref{eq:omega-root}) in ascending powers of $\epsilon^{-1}$,
i.e. 
\begin{equation}
\mathrm{Im}(\omega')\sim\frac{1}{2\epsilon}+O\left(\frac{1}{\epsilon^{2}}\right).\label{eq:damping-high-viscosity}
\end{equation}
This agrees with the asymptotic approximations by \citet{Levich1962} which suggests that increasing viscosity would decrease damping for $\epsilon\gg\epsilon^{\star}$. 

Using Eq.\,(\ref{eq:CRIT-ANALYTIC}), the analytical critical wavelength (the value of $\lambda$ with associated damping ratio $\zeta=1$) of the capillary wave is given by 
\begin{equation}
\lambda_c^\star = \frac{2\pi}{\epsilon^{\star2}}l_\mathrm{vc}.\label{eq:crit-wavelength}
\end{equation}
For water under rtp., we have $\lambda_c^\star\doteq 40.1894\mathrm{nm}$. In comparison, a harmonic oscillator approximation \citep{Denner2016b} in Eq.\,(\ref{eq:HARMONIC-Crit}) gives the result $\lambda_c^{(0)}\doteq40.9838\mathrm{nm}.$ \changes{Consider the relative error between the harmonic oscillator and the analytical critical wavelengths  
\begin{equation}
\left|\frac{2^{1/3}}{\Theta}-\frac{2}{\epsilon^{\star2}}\right|\frac{\epsilon^{\star2}}{2}\simeq0.01977,	
\end{equation}
we observe the system is largely second-order in the neighbourhood of the critical wavelength, as the harmonic oscillator value of $\lambda_c$ is within 2\% of the analytical value from the wave dispersion.}

\section{Damping ratio for a generalised system}

\changes{For systems of a higher order, the damping ratio $\zeta$ is not naturally defined and we usually inspect the root-locus diagram in order to decompose the system into a sum of first- and second-ordered systems to provide an estimate calculation. Here we consider a numerical method to obtain an equivalent damping ratio, whereby $\zeta\geqslant1$ when the area of the amplitude below the settling value vanishes for all time almost everywhere. The critical wavelength is then the supremum of the set of wavelengths such that the above property holds. We express this as 
\begin{equation}
\lambda_{c}=\sup_{\lambda\in\mathbb{R}^{+}}\left\{ \lambda:\lim_{T\rightarrow\infty}\int_{0}^{T}\mathrm{A}(\lambda,\tau)\left[1-H(\mathrm{A}(\lambda,\tau))\right]\,\mathrm{d}\tau\rightarrow0\,\mathrm{a.e.}\right\}, \label{eq:area-method}
\end{equation}
where $H(x)$ is the Heaviside step function. 

For underdamped waves, even for second-order systems, the logarithmic decrement or fractional overshoot methods tend to break down or become less accurate near regions of critical damping.  Hence, to determine the damping ratios
in the neighbourhood of $\zeta\approx1$, we adapt the area
method in Eq.\,(\ref{eq:area-method}). Consider that the area under the $t$-axis is given by the function
\begin{equation}
\Xi(\zeta)=\int_{0}^{\infty}\mathrm{d}t\,\left\{ \mathrm{\Lambda}(\zeta,t)\,\left[1-H(\mathrm{\Lambda}(\zeta,t))\right]\right\} 
\end{equation}
where $\Lambda(\zeta,t)$ satisfies the normalised harmonic oscillator
equation 
\begin{equation}
\frac{\mathrm{d}^{2}\Lambda}{\mathrm{d}t^{2}}+2\zeta\frac{\mathrm{d}\Lambda}{\mathrm{d}t}+\Lambda=0\quad\mathrm{subject\ to\ }\Lambda(\zeta,0)=1,\ \frac{\mathrm{d}\Lambda}{\mathrm{d}t}(\zeta,0)=0.
\end{equation}
The generalised (numerical) damping ratio for $\zeta\leqslant1$ can
then be obtained by the inverse operation
\begin{equation}
\zeta=\Xi^{-1}(X),
\end{equation}
where $X$ is the area under the $t$-axis of a generalised system. This numerical method agrees well with logarithmic decrement and fractional overshoot schemes in the relevant underdamped regimes. 

In cases where the higher order system can readily be approximated locally by a second-order system, we note that its dominant poles have a larger residue and time constant $t_c=1/(\zeta_n\omega_n)$ relative to other poles, where $\zeta_n$ and $\omega_n$ are the damping ratio and frequency associated with each pole. In cases that are not clear cut, i.e. where all the poles are closer together with $t_c$ and residues of a similar magnitude, the numerical definition of the damping ratio above only provides an estimate of the true damping ratio and we need to examine the poles in more detail. 

To encode such information into a convenient form, we construct
the \textit{minimal pole matrix} $\{\zeta(\lambda)\}$. To decompose the system, we first consider the minimal realisation of the transfer function. For a general system, let $\Theta_{1}=\{q_{i}\in\mathbb{C}:\ \mathrm{Q}(-q_{i})=0\}$
and $\Theta_{2}=\{p_{j}\in\mathbb{C}:\mathrm{P}(-p_{j})=0\}$ be the
set of poles and zeros of the transfer function $\mathrm{tf}(s)\equiv\mathrm{P}(s)/\mathrm{Q}(s)$,
which can be written in the form, 
\begin{equation}
\mathrm{tf}(s) = \prod_{q_{i}\in\Theta_{1},p_{j}\in\Theta_{2}}\left(\frac{s+p_{j}}{s+q_{i}}\right).
\end{equation}
% & =& \frac{a_{0}}{s}+\sum_{i=1}^{|\Theta_{1}^{(1)}|}\frac{a_{i}}{s+q_{i}}+\sum_{j=1}^{|\Theta_{1}^{(2)}|}\frac{b_{j}(s+\zeta_{j}\omega_{j})+c_{j}\omega_{j}(1-\zeta_{j}^{2})^{1/2}}{s^{2}+2\zeta_{j}\omega_{j}s+\omega_{j}^{2}}, 
%where $a_i,b_j,c_j\in\mathbb{C}$, $\Theta_{1}^{(1)}=\Theta_{1}\cap\mathbb{R}$ and $\Theta_{1}^{(1)}\oplus\Theta_{1}^{(2)}=\Theta_{1}.$ 
Applying the pole-zero cancellation procedure, we obtain the minimal realisation of the
transfer function, henceforth known as the \textit{minimal transfer function} \textit{$\mathrm{mtf}(s)$}, defined by 
\begin{eqnarray}
\mathrm{mtf}(s,\vartheta) & = & \frac{\mathrm{P}_{m}(s)}{\mathrm{Q}_{m}(s)}\\
 & = & \sum_{q_{i,m}\in\Theta_{1,m}(\vartheta)}\frac{\mathrm{res}(-q_{i,m})}{s+q_{i,m}}
\end{eqnarray}
where the polynomials $\mathrm{P}_m(s)$ and $\mathrm{Q}_m(s)$ satisfy
\begin{equation}
	\deg(\mathrm{Q}_{\mathrm{m}}+\mathrm{P_{m})\leqslant\deg(\mathrm{P}+\mathrm{Q})}
\end{equation}
and $\Theta_{1,m}(\vartheta) \subseteq\Theta_{1}$ is the set of
poles of $\mathrm{Q}$ with significant residues (tolerance of order $\vartheta$) 
\begin{equation}
	\Theta_{1,m}(\vartheta)=\left\{ q_{i,m}\in\mathbb{C}:\ \mathrm{Q}(-q_{i,m})=0,\ \frac{\left|\mathrm{res}(q_{j})\right|}{\max_{q_{i}\in\Theta_{1}}\left|\mathrm{res}(q_{i})\right|}= O(\vartheta)\right\}.
\end{equation}

Returning to the construction of the minimal pole matrix, we let the first column of the matrix illustrate the 
order of the poles of the minimal transfer function in dots form; the second column considers the relative magnitudes of
their time constant $t_c$; the third column compares their relative residues; the 
fourth column gives the damping ratios $\zeta_n$ associated with each pole.
Furthermore, to the right of the line separator, we provide an estimated equivalent
second-order damping ratio of the entire system using the area numerical
method described previously. We say that a system is second-order dominant if one set of complex conjugate poles dominate the other poles (i.e. having the largest $t_c$ and residue). In diagrammatic form for $\vartheta=1$, we have 
\begin{eqnarray}
	\{\zeta(\lambda;\beta,\mathrm{B})\}=\left\{ \left.
	\begin{array}{cccc}
\mathrm{pole} & \mathrm{relative} & \mathrm{relative} & \mathrm{associated }\\
 \mathrm{type} & t_c & \mathrm{residue} & \zeta_n \\
\{\bullet,\bullet\bullet\} & (0,10) & (0,10) & [0,\infty)
\end{array}\right|
\begin{array}{c}
\mathrm{numerical }\\
\mathrm{\zeta}\\{}
[0,\infty)
\end{array}\right\}
\end{eqnarray}
For example, for $\lambda=6.22\lambda_c^\star$ at rtp., the clean case for water exhibits the following minimal pole matrix 
\begin{eqnarray}
{\scriptstyle \left\{ \left.\begin{array}{cccc}
\bullet\bullet & 1 & 1 & 0.75\\
\bullet\bullet & 1 & 2.30 & 0.45
\end{array}\right|{\textstyle 0.45}\right\} },
\end{eqnarray}
from which we observe that the system is second-order dominant since the set of complex conjugate poles with associate damping ratio 0.45 dominates (in the sense of residue) the other and, thus, we can deduce that the true damping ratio of the system is close to the approximate second-order value.

We take this analysis to the region near critical damping, i.e. for $\zeta\rightarrow1^{+}$ and $\zeta\rightarrow1^{-}$. In the clean case, we take the analytical result $\lambda_{c}^{\star}=2\pi l_{\mathrm{vc}}/\epsilon^{\star2}$
to be the definition of the critical (damping) wavelength. The relevant minimal pole matrices take the form 
\begin{eqnarray}
\{\zeta(\lambda\rightarrow\lambda_{c}^{\star+})\} & ={\scriptstyle \left\{ \begin{array}{cccc}
\bullet\bullet & 1 & 1 & 1\\
\bullet\bullet & \ \,1^{-} & \ \,1^{+} & < 1
\end{array}\right\} },\\
\{\zeta(\lambda\rightarrow\lambda_{c}^{\star-})\} & ={\scriptstyle \left\{ \begin{array}{cccc}
\bullet & \ \, 1^{-} & 1 & 1\\
\bullet & 1 & \ \,1^{-} & 1
\end{array}\right\} .}
\end{eqnarray}
We can see that this transition from $\lambda_{c}^{\star+}\rightarrow\lambda_{c}^{\star-}$
for the clean case boasts a transformation of the complex conjugate
into two separate first-order poles. Or in diagrammatic form, we have 
\begin{equation}
\begin{array}{c}
\bullet\bullet\\
\bullet\bullet
\end{array}\rightarrow\begin{array}{c}
\bullet\\
\bullet\\
\end{array},
\end{equation}
i.e. a $2^{2}$ to $1^{2}$ transition.  

Extending to contaminated
systems, we summarise the results of the minimal pole matrices of the system at the critical transition in table 1, \cchanges{where the notation $1^a2^b$ denotes a system with $a$ first-order and $b$ second-order poles with significant relative residue. }
\begin{table}
\begin{centering}
\begin{tabular}{cccccccc}
 & $\beta=\mathrm{B}=0$ &  & $\beta>0,\,\mathrm{B}=0$ &  & $\beta=0,\,\mathrm{B}>0$ &  & $\beta,\mathrm{B}>0$\tabularnewline
 & $\begin{array}{c}
\bullet\bullet\\
\bullet\bullet
\end{array}\rightarrow\begin{array}{c}
\bullet\\
\bullet
\end{array}$ &  & $\begin{array}{c}
\bullet\bullet\\
\bullet\bullet
\end{array}\rightarrow\begin{array}{c}
\bullet\\
\bullet
\end{array}$ &  & $\begin{array}{c}
\bullet\\
\bullet\bullet\\
\bullet\bullet
\end{array}\rightarrow\begin{array}{c}
\bullet\\
\bullet
\end{array}$ &  & $\begin{array}{c}
\bullet\\
\bullet\bullet\\
\bullet\bullet
\end{array}\rightarrow\begin{array}{c}
\bullet\\
\bullet
\end{array}$\tabularnewline
 & $2^{2}\rightarrow1^{2}$ &  & $2^{2}\rightarrow1^{2}$ &  & $1^{1}2^{2}\rightarrow1^{2}$ &  & $1^{1}2^{2}\rightarrow1^{2}$\tabularnewline
\end{tabular}
\par\end{centering}
\caption{Schematic of poles at the transition from $\lambda_{c}^{\star+}\rightarrow\lambda_{c}^{\star-}$ for water at room temperature and pressure.}
\end{table}
We note that the effect of surface viscosity is to introduce an extra first-order pole (with unit damping ratio) to the system, while the Marangoni effect does not change the pole composition for the underdamped region before the critical damping transition. Moreover, we observe at this critical damping transition that the system enters overdamped regime if its minimal pole matrix is of the $1^2$-type irrespective of its type in the underdamped regime prior to the transition.  Henceforth, we shall define a generalised higher-order (capillary) wave to be in overall overdamped motion if its minimal pole matrix is of the $1^2$-type. 

Using the definition of the overdamped regime for a generalised higher-order system, we determine numerically the corrections to the critical wavelength for increasing surface viscosity (through $\mathrm{B}$), surfactant concentration (through $\beta$) and bulk viscosity \cchanges{through $\theta$, where 
\begin{equation}
\nu=(1+\theta)\nu_0
\end{equation}}in figure 3.  We note that the curves denoting surface viscosity and the Marangoni effect intersect near $\mathrm{B}\simeq 0.65$ and that while the Marangoni curve is roughly exponential, the surface viscosity curve is the sum of two exponential functions. Moreover, we observe that a small amount of surface viscosity present in the system has an amplified effect on the system and that a 7-fold increase in critical wavelength for $\mathrm{B}=1$ results in very different dynamics and mechanisms as the sub-100nm brings forward the possibility of long-range molecular interactions as well as the hydrodynamics. Also of consideration is the proximity of the critical wavelength to the wavelengths of the visible spectrum of light which allows thin film behaviours to be captured by light scattering methods.  Hence the presence of surfactants could determine whether or not we would be within such a range to allow interferometry techniques. 

\begin{figure}
\begin{center}
	\includegraphics[width=13cm]{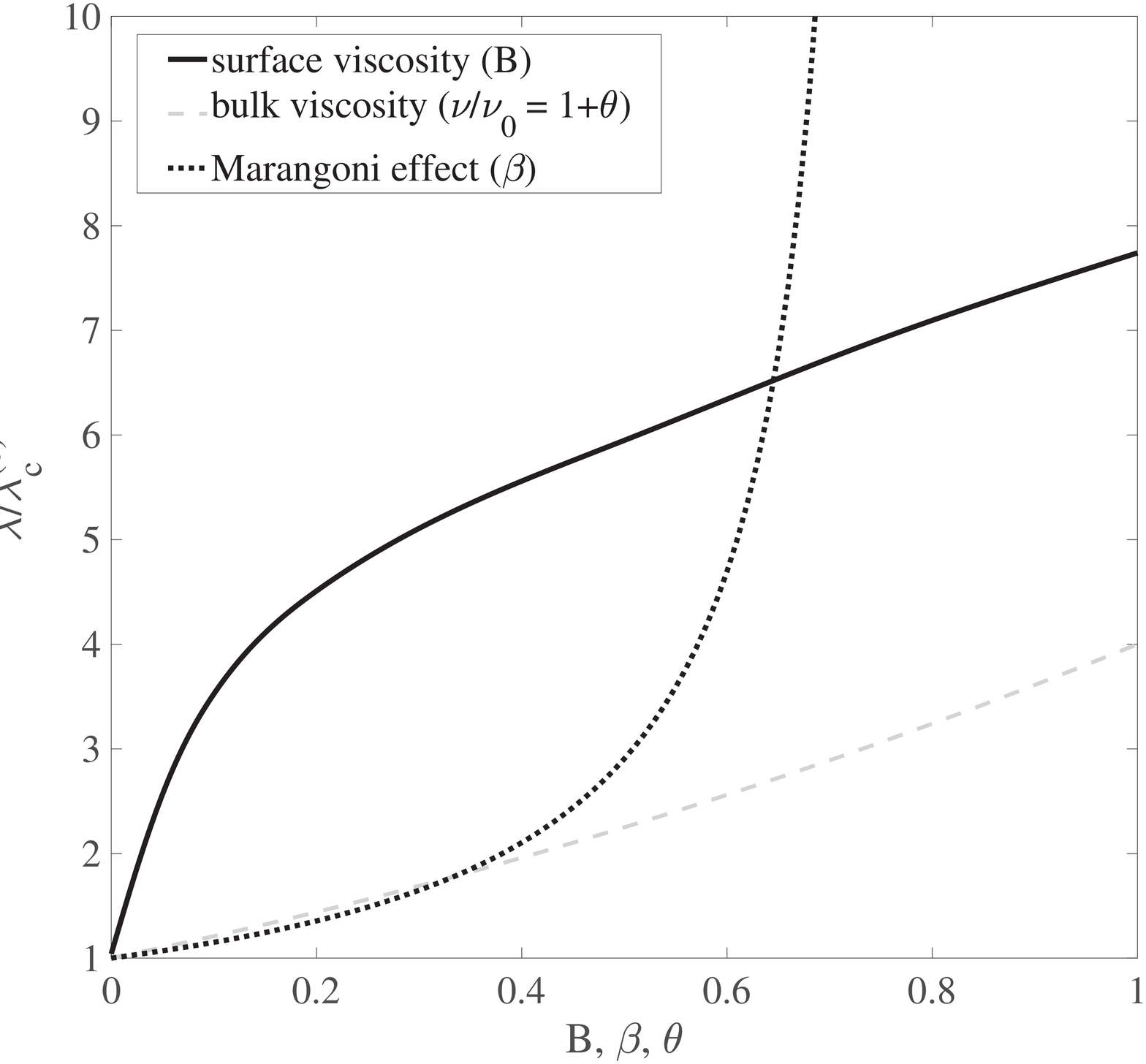}
\end{center}
\caption{\cchanges{Independent} corrections to the critical wavelength for increasing surface viscosity (via $\mathrm{B}$, \cchanges{with $\theta=\beta=0$}), bulk viscosity (via $\theta$, where $\nu/\nu_0=1+\theta$ \cchanges{and $\beta=\mathrm{B}=0$}) and Marangoni effect (through $\beta$\cchanges{, with $\theta=\mathrm{B}=0$}). }
\end{figure}

Comparisons of the range under  with experimental and computational results on surface viscosity can be made using values reported previously in the literature. Experimentally, \citet{Kanner1969} summarises in table 2 the surface viscosities for both surfactant and polymeric films, where, for a dilute amount of sodium lauryl sulfate and polydimethylsiloxane in particular, the surface viscosity corresponds to a lower bound of $\mathrm{B}=O(1)$ for $k= O(10^6)$. A similar correspondence can be found with the upper bound surface \cchanges{shear} viscosity of $O(10^{-8}\mathrm{Nsm^{-1}})$ found in \citep{Zell2014} for soluble surfactants. \cchanges{We note that this measurement does not include surface dilatational viscosity and so corresponds to the case \citep{Lucassen1966} where diffusional transport between the surface and the bulk is neglected, and assumes that the bulk viscosity and density are constant right up to the interface.} More recently, \citet{Gounley2016} characterises the influence of \cchanges{both shear and dilatational} surface viscosity on droplets in shear flow in the range $\mathrm{B}=O(10^0)$ to $O(10^1)$ for a range of capillary numbers.

 Beyond the cmc value, the Marangoni effect should in principle have no overall contribution to the capillary wave; the dotted curve in figure 3 would end abruptly at the cmc value. The combination of surface viscosity together with the Marangoni effect is however not straightforward; as in previous experimental studies \citep{Brown1953, Kanner1969}, surface viscosity also appears to alter the ability of the Marangoni effect to lower surface tension. It would therefore be fruitful in a future contribution to investigate this surfactant interference mechanism through a more systematic experimental and theoretical study. In particular, a numerical approach similar to that of \citet{Sinclair2018} could include the usage of a nonlinear equation of state for the surface tension coefficient $\sigma$. The effect of the deviation from the linear equation of state on the amplitude of the capillary wave would aid the analysis near the cmc value of the surfactant solution. 
 
 %For instance, an increase of 30\% in bulk viscosity is comparative to a cmc value that corresponds to a roughly 38\% maximum reduction to the initial surface tension $\sigma_0$ by the Marangoni effect. In reality, the $\beta$-curve in figure 3 would end abruptly at the point which corresponds to the cmc and level off instead of increasing exponentially. Beyond the cmc value, the Marangoni effect should in principle have no overall contribution to the capillary wave and one would need to increase bulk viscosity to increase the critical wavelength. It is also the case that the linear equation of state would not be an accurate description near the cmc and more work is needed near and above that region. The further addition of surface viscosity together with the Marangoni effect is however not straightforward; as in previous experimental studies \citep{Brown1953, Kanner1969}, surface viscosity also appears to alter the ability of the Marangoni effect to lower surface tension. It would therefore be fruitful in a future contribution to investigate this surfactant interference mechanism through a more systematic experimental and theoretical study. 
}

\section{Conclusion}

In this work the surface viscosity effect has been incorporated into the  integro-differential initial value problem describing the
wave dynamics of small-amplitude capillary waves via the Boussinesq-Scriven surface model. We have shown that, particularly at lengthscales close to the critical damping wavelength, a very small amount of surface viscosity can  dramatically increase the critical wavelength of the capillary waves, \cchanges{in contrast with the Marangoni effect which becomes prominent at larger wavelengths.} In view of the important role that capillary waves play in inducing the rupture process of thin films \citep{Aarts2004}, we anticipate the various interfacial phenomena controlling the wave dynamics at the very minute lengthscale to contribute towards the understanding of the stability of foams with non-trivial surface viscosity. \changes{In particular, the correction of the critical wavelength due to surface viscosity and Marangoni effects, which we summarised in figure 3 using numerical methods, is bound to alter the onset of fluid instabilities for very thin liquid films. It is also useful towards the optimisation of additives to achieve the desired increases in the critical wavelength. Finally, we expect the concept of a critical damping wavelength and its correction by surface material to be useful to a further number of general interfacial phenomena, such as the onset of thin film quasi-elastic wrinkling and Faraday-like instabilities in the same lengthscale.}

\begin{acknowledgements}
The authors acknowledge the financial support of the Shell University Technology Centre for fuels and lubricants and the Engineering and Physical Sciences Research Council (EPSRC) through grants EP/M021556/1 and EP/N025954/1. 
\end{acknowledgements}

\appendix
\changes{\section{$\deg\mathrm{Q}-\deg \mathrm{P}\geqslant2 \,\Rightarrow\, \mathrm{Z}(n,0)=0$}
Consider the rational expression 
\begin{eqnarray}
\hat{\mathrm{f}}(s)  &\equiv &\frac{\mathrm{P}(s,m)}{\mathrm{Q}(s,n)}\\
  &=& \frac{s^{m}+\varpi_{1}s^{m-1}+\cdots+\varpi_{m-1}s+\varpi_{m}}{s^{n}+\varsigma{}_{1}s^{n-1}+\cdots+\varsigma_{n-1}s+\varsigma_{n}} \label{eq:RATEXP}
\end{eqnarray}
where $\mathrm{P}(s,m)$ is a polynomial of order $m$ in $s$, 
\begin{equation}
\mathrm{Q}(s,n)=\prod_{i=1}^{n}(s-q_{i})
\end{equation}
 is a polynomial of order $n>m$ in $s$ with distinct roots $q_{i}$
and 
\begin{equation}
\sum_{1\leqslant i_{1}<i_{2}<\cdots<i_{k}\leqslant n}q_{i_{1}}q_{i_{2}}\cdots q_{i_{k}}=(-1)^{k}\varsigma_{n-k}.
\end{equation}
Rewriting $\hat{\mathrm{f}}(s)$ using a partial fraction decomposition,
we have 
\begin{equation}
\hat{\mathrm{f}}(s)=\sum_{i=1}^{n}\frac{\mathrm{P}(q_{i})}{\mathrm{Q}'(q_{i})}\frac{1}{s-q_{i}}
\end{equation}
and taking an inverse Laplace transform gives 
\begin{eqnarray}
\mathrm{f}(t) & = &\sum_{i=1}^{n}\frac{\mathrm{P}(q_{i})}{\sigma_{i}^{(n)}(q_{i})}\mathrm{e}^{-q_{i}t}\\
 & =& \sum_{j=0}^{\infty}(-1)^{j}\mathrm{Z}(n,j)\frac{t^{j}}{j!},
\end{eqnarray}
where $q_{i}$ are roots of the polynomial $\mathrm{Q}(s,n)$ and
\begin{equation}
\mathrm{Z}(n,j)=\sum_{k=1}^{n}\frac{\mathrm{P}(q_{j})}{\sigma_{j}^{(n)}(q_{j})}q_{k}^{j}.\label{eq:Z(NJ)}
\end{equation}
Expansion of Eq.$\,$(\ref{eq:RATEXP}) for large $s$ and inversion
term-wise gives 
\begin{equation}
\mathrm{f}(t)\sim\frac{t^{n-m-1}}{(n-m-1)!}+\frac{(\varpi_{1}-\varsigma_{1}\varpi_{m})t^{n-m}}{(n-m)!}+O(t^{n-m+1}).
\end{equation}
Comparing with Eq.$\,$(\ref{eq:Z(NJ)}) shows that $\mathrm{Z}(n,j)=0$
 if 
\begin{equation}
 	0\leqslant j\leqslant n-m-2,
\end{equation}
which reduces to the condition
\begin{equation}
\deg\mathrm{Q}-\deg\mathrm{P}\geqslant2.
\end{equation}

\section{The contaminated wave dispersion relation}
Following \citet{Lamb1932} and using the equations of motion in \S3, the dispersion relation for a contaminated surface with non-trivial Marangoni effect and surface viscosity can be obtained from the determinant of the matrix $M$, given by 
\begin{equation}
M=\begin{pmatrix}
\dfrac{1}{n}\left(n'^{2}+2\epsilon n'+1\right) & \dfrac{\mathrm{i}}{n}\left[1+2\epsilon n'\left(1+\dfrac{n'}{\epsilon}\right)^{1/2}\right]\\
\mathrm{i}\left(\dfrac{\beta}{n'}+(2+\mathrm{B})\epsilon\right) &  n'+2\epsilon+\left(\dfrac{\beta}{n'}+\mathrm{B}\epsilon\right)\left(1+\dfrac{n'}{\epsilon}\right)^{1/2}
\end{pmatrix},
\end{equation}
where we considered the wave-form solution 
\begin{equation}
\begin{pmatrix}u\\
v\\
F
\end{pmatrix}=\begin{pmatrix}-\mathrm{i}kA\mathrm{e}^{ky}-mC\mathrm{e}^{my}\\
-kA\mathrm{e}^{ky}+\mathrm{i}kC\mathrm{e}^{my}\\
-\dfrac{k}{n}(A-\mathrm{i}C)
\end{pmatrix}\exp\left(\mathrm{i}kx+nt\right)
\end{equation}
for the fluid velocities $u,v$ and the free surface $F$, where $A,C\in \mathbb{C}$, $n=\mathrm{i}\omega$ and $n'=\mathrm{i}\omega/\omega_0$ for
\begin{equation}
m^{2}=k^{2}+\frac{n}{\nu}
\end{equation}
and 
\begin{equation}
\omega_0^2=\mathrm{g}k+\frac{\sigma_0 k^3}{\rho}.	
\end{equation}
Similarly, the pressure $p$ is given by 
\begin{equation}
	\frac{p}{\rho}=An\exp\left(ky+\mathrm{i}kx+nt\right)-\mathrm{g}y.
\end{equation}
Evaluating the determinant of $M$ gives
\begin{eqnarray}
\left(Z^{4}-1+\frac{1}{\epsilon^{2}}\right)\left[1+Z^{2}+\left(\frac{\beta}{\epsilon^{2}(Z^{2}-1)}+\mathrm{B}\right)Z\right]\nonumber \\
=\left(\frac{1}{\epsilon^{2}}+2(Z^{2}-1)Z\right) &  & \left[2+\mathrm{B}+\frac{\beta}{\epsilon^{2}(Z^{2}-1)}\right]
\end{eqnarray}
where
\begin{equation}
\frac{n'}{\epsilon}=Z^{2}-1.
\end{equation}
Factorising yields
\begin{eqnarray}
W_0(Z;\epsilon)\left(1+Z+\mathrm{B}+\frac{\beta}{\epsilon^{2}(Z^{2}-1)}\right)\nonumber \\
=-\left(\frac{\beta}{\epsilon^{2}(Z^{2}-1)}+\mathrm{B}\right) & (Z-1)^{3}.&\label{eq:fullwavedisper}
\end{eqnarray}
where $W_0(Z;\epsilon)=0$ is the dispersion relation clean case result. This can be shown if we let $\beta,\mathrm{B}=0$ in Eq.\,(\ref{eq:fullwavedisper}), i.e. 
\begin{equation}
\left(Z^{4}-1+\frac{1}{\epsilon^{2}}\right)\left(Z^{2}+1\right)=2\left(\frac{1}{\epsilon^{2}}+2(Z^{2}-1)Z\right).
\end{equation}
Factorising gives 
\begin{equation}
\left(Z^{4}+2Z^{2}-4Z+1+\frac{1}{\epsilon^{2}}\right)(Z+1)=0,
\end{equation}
which reduces to 
\begin{equation}
	W_0(Z;\epsilon)=0
\end{equation}} 
if we neglect the spurious root $Z+1=0$.
\bibliographystyle{jfm}
\bibliography{Bib.bib}
\end{document}